\documentclass{aa}
\usepackage{float}
\usepackage{graphicx}
\usepackage{txfonts}
\usepackage{hyperref}
\usepackage{multirow}
\usepackage{lscape}
\usepackage{amsmath}
\graphicspath{{Main_figure/}}

\begin{document}

\title{The robustness in identifying and quantifying high-redshift bars using JWST observations}

   \author{Xinyue Liang\inst{\ref{XMU}},
   Si-Yue Yu\inst{\ref{mpifr},}\inst{\ref{ipmu}},
   Taotao Fang\inst{\ref{XMU}},
   \and
   Luis C. Ho\inst{\ref{KIAA},}\inst{\ref{PKU}}
          }

   \institute{ 
   Department of Astronomy, Xiamen University, Xiamen, Fujian 361005, People's Republic of China \label{XMU}
   \and
    Max-Planck-Institut für Radioastronomie, Auf dem Hügel 69, 53121 Bonn, Germany  \\ \email{phyyueyu@gmail.com, si-yue.yu@ipmu.jp}\label{mpifr} 
    \and
    Kavli Institute for the Physics and Mathematics of the Universe (WPI),The University of Tokyo Institutes for Advanced Study, The University of Tokyo, Kashiwa, Chiba 277-8583, Japan \label{ipmu}
    \and
   The Kavli Institute for Astronomy and Astrophysics, Peking University, 5 Yiheyuan Road, Haidian District, Beijing, 100871, China\label{KIAA}
   \and
   Department of Astronomy, Peking University, 5 Yiheyuan Road, Haidian District, Beijing, 100871, China \label{PKU}
            }

\abstract{
Understanding the methodological robustness in identifying and quantifying high-redshift bars is essential for studying their evolution with the {\it James} {\it Webb} Space Telescope (JWST). We used nearby spiral galaxies to generate simulated images at various resolutions and signal-to-noise ratios, and obtained the simulated galaxy images observed in the Cosmic Evolution Early Release Science (CEERS) survey from Yu et al. Through a comparison of measurements before and after image degradation, we show that the bar measurements for massive galaxies remain robust against noise.  While the measurement of the bar position angle remains unaffected by resolution, the measured bar ellipticity is significantly underestimated in low-resolution images. The size measurement is on average barely affected as long as the intrinsic bar size $a_{\rm bar,\,true}>2\times{\rm FWHM}$. To address these effects, correction functions are derived. We also find that the effectiveness of detecting bars remains at $\sim$\,100\% when the $a_{\rm bar,\,true}/{\rm FWHM}$ is above 2, below which the rate drops sharply, quantitatively validating the effectiveness of using $a_{\rm bar,\,true}>2\times {\rm FWHM}$ as a bar detection threshold.  
We analyze a set of simulated CEERS images, which take into account observational effects and plausible galaxy (and bar-size) evolution models, and show that a significant (and misleading) reduction in detected bar fraction with increasing redshift would apparently result even if the true bar fraction remained constant.
Our results underscore the importance of disentangling the true bar fraction evolution from resolution effects and bar size growth. 
}

\maketitle

\section{Introduction}

A galactic bar is a linear elongated stellar structure spanning the center of a disk galaxy. In the local Universe, there are $\sim$\,70\% of disk galaxies hosting a bar when viewed in optical or near-infrared (NIR) wavelengths \citep[e.g.][]{Mendez2007,MarinovaandJogee2007,Aguerri2009, Ho2011,Buta2015,Erwin2018, Yu2022b}. The bar fraction may vary with Hubble types, stellar mass, and color index \citep{NairandAbraham2010,Barazza2008,Diaz-Garcia2016, Erwin2018}. It is generally accepted that stellar bars play an important role in galaxy evolution.  The non-axisymmetric bar gravitational potential drives cold gas flow toward the galaxy center along the bar's dust lane, enhancing central star formation and leading to the growth of pseudo bulges \citep[][]{Athanassoula1992, Athanassoul2002, AthanassoulaLambertDehnen2005, KormendyKennicutt2004, AthanassoulaLambertDehnen2005, Jogee2005, Ellison2011, Wang2012, Wang2020, Gadotti2020, Yu2022a, Yu2022b}.  Meanwhile, bars reshape galaxy morphology by rearranging the mass distribution, forming substructures such as disk break, spiral arms, and rings \citep[e.g.,][]{Knapen1995,KormendyKennicutt2004,Ellison2011, ErwinDebattista2013, Gadotti2020}.

Through observations from the {\it Hubble} Space Telescope (HST), the bar fraction is found to evolve with redshift. Early HST study by \cite{Abraham1999} reported a decline in the bar fraction towards higher redshifts \citep[see also][]{Abraham1996, VandenBergh1996}, while later \cite{Elmegreen2004} and \cite{Jogee2004} found that the bar fraction remains consistent up to $z\approx1$. With the dataset of a considerably large sample covering a wide mass range, \citet{Sheth2008} demonstrated that the bar fraction decreases from 65\% at $z=0.2$ to 20\% at $z=0.84$ \citep[see also][]{Cameron2010}. The declining trend was further confirmed by \cite{Melvin2014} using visual classifications from the Galaxy Zoo \citep{Willett2013}.  By studying barred galaxies at $0.2<z\leq 0.835$, \cite{Kim2021} found that  normalized bar sizes do not exhibit any clear cosmic evolution, implying that bar and disk evolution are closely intertwined throughout time. Simulation work by \cite{Kraljic2012} showed that the bar fraction drops to nearly zero at $z \approx 1$, suggesting bars providing as a tool to identify the transition epoch between the high-redshift merger-dominated or turbulence-dominated disks and local dynamically settled disks. Nevertheless, a recent study on the IllustrisTNG galaxies found that the bars appear as early as at $z = 4$ and the bar fraction evolves mildly with cosmic time \citep{Rosas-Guevara2022}. They argued that if only considering long bars their results can reconcile with the observation studies as those can suffer from resolution effects.

Analysing bar structures in galaxies at high redshift using HST observations present a considerable challenge. For galaxies at $z\gtrsim2$, imaging through the HST F814W filter observes the rest-frame ultraviolet, a wavelength where bars are often less visible \citep{Sheth2003}. The HST NIR F160W imaging possesses a relatively broader point spread function (PSF), making it inadequate for resolving bar structures at high redshift. In addition, the depth of HST observations might be insufficient for detecting the outer regions of high-redshift galaxies, causing galaxies with a long bar to resemble edge-on galaxies. The {\it James Webb} Space Telescope (JWST) delivers images with unparalleled sensitivity and resolution in NIR, significantly enhancing our understanding of galaxy structures at high redshift. Recent JWST studies have revealed a significant fraction of regular disk galaxies at high redshift \citep{Ferreira2022a, Ferreira2022b, Kartaltepe2023, Nelson2022, Robertson2023, Jacobs2023, XuYu2024}. This result contrasts with the findings from HST-based studies, which predominantly identified peculiar galaxies at $z>2$ \citep[e.g.,][]{Conselice2008, Mortlock2013}.  Remarkably, barred galaxies at $z\approx2$\textendash3, previously undetected in HST observations, have now been identified using JWST \citep{Guo2023, LeConte2024, Costantin2023}.

Although JWST imaging offers exceptional angular resolution, when observing high-redshift galaxies, its physical resolution is still lower than that achieved with ground-based observations of nearby galaxies. This complicates direct comparisons between bars observed in low-redshift and high-redshift galaxies, making the study of bar evolution less straightforward. \cite{Sheth2003} pointed out that the bar fraction calculated using low-resolution images tends to be underestimated. This underestimation is likely to be exacerbated by the bar size evolution, wherein bars become shorter in physical size at higher redshift as predicted in numerical simulations \citep[e.g.,][]{Debattista2000, Martinez2006, Athanassoula2013, Algorry2017}. In contrast, the influence of cosmological surface brightness dimming on bar detection is minimal \citep{Sheth2008}.  By analysing apparent the bar sizes in galaxies from the Spitzer Survey of Stellar Structure in Galaxies \citep[S$^4$G;][]{Sheth2010}, \cite{Erwin2018} showed that most of the projected bar radii are larger than twice the PSF full width at half-maximum (FWHM) and thus suggested this value as the bar detection threshold. Such a threshold successfully reconciles the difference in the dependency of bar fraction on parameters like stellar mass or gas fraction, especially when comparing their results with studies based on low-resolution SDSS images \citep{Masters2012, Oh2012, Melvin2014, Gavazzi2015}.  Likewise, based on extensive experiments on artificial galaxies, \cite{Aguerri2009} suggested that the limitation of bar detection is 2.5 times FWHM.
Nevertheless, the threshold bar size for bar detection have not yet been validated in a quantitative way based on real images, especially under typical JWST observational conditions. Resolution may also impact measured bar properties, such as size, ellipticity, and position angle, which are frequently employed to study the formation and evolution of bars \citep[e.g.,][]{Elmegreen2007, Gadotti2011, Kim2021,Yu2022b}.  While it seems intuitive that bar ellipticity would be underestimated due to PSF smoothing as spatial resolution deteriorates, the exact impact and the influence on other parameters remain largely unexplored.

It is therefore nontrivial to interpret the observational results obtained from JWST without knowing the systematics caused by the observation limitations. To disentangle the potential intrinsic relations or evolution for bars from the observation effects, in the current work we aim to understand how the observational factors can influence the identification and quantification of bars. Recently, by accounting for observational effects and galaxy evolution, \cite{Yu2023} used a sample of nearby galaxies to create images of simulated high-redshift galaxies as would be observed by JWST in the Cosmic Evolution Early Release Science (CEERS) survey (PI: Finkelstein, ID=1345, \citealt{Finkelstein2022a}; \citealt{Bagley2023}).  In addition to the simulated galaxies CEERS images provided by \cite{Yu2023}, we employ their sample of nearby galaxies to produce images at low resolutions for a given $S/N$ and at low $S/N$ for a given resolution. Then we compare the measurements before and after image degradation to understand the robustness in analysing bars in high-redshift galaxies observed with JWST. The structure of this paper is as follows. Section~\ref{sect:obs} presents an overview of the dataset. Section~\ref{sect:sim} describes the procedure for generating images mimicking the JWST resolution and S$/$N. In Section~\ref{sect:results}, we present the robustness of bar structure measurements under JWST observation. We discuss the implication of our results in Section~\ref{sect:implication}. A summary is given in Section~\ref{sect:conclusions}. Throughout this work, we use AB magnitudes and assume the following cosmological parameters: $(\Omega_{\rm M}, \Omega_{\Lambda}, h)=(0.27, 0.73, 0.70)$.

\section{Observational material}\label{sect:obs}
We construct our sample to study the robustness in identifying and quantifying bars based on the nearby galaxies sample defined in \cite{Yu2023}. By restricting to luminosity distance ($D_L$) of $12.88 \leq D_L \leq 65.01$\,Mpc, stellar mass ($M_{\star}$) of $10^{9.75 \text{\textendash} 11.25} M_{\odot}$, and excluding images severely contaminated by close sources, \cite{Yu2023} select 1816 galaxies from the Siena Galaxy Atlas \footnote{https://www.legacysurvey.org/sga/sga2020/}\citep[SGA;][]{Moustakas2021} that is made up of 383,620 galaxies from the Dark Energy Spectroscopic Instrument (DESI) Legacy Imaging Surveys \citep{Dey2019}. Out of the 1816 galaxies, we select 448 face-on spiral galaxies by requiring that the galaxies are available in the third Reference Catalog of Bright Galaxies \citep[RC3;][]{vaucouleurs1991}, and have Hubble type of {\it T} $ > 0.5$ and axis ratio of {\it b$/$a} $>$ 0.5. We exclude S0s because they are much rarer at high redshifts than in the local universe \citep[e.g.,][]{Postman2005, Desai2007, Cavanagh2023}, their bars are significantly different from those in spirals \citep[e.g.,][]{Aguerri2009,Buta2010,Diaz-Garcia2016}, and distinguishing them from Es is challenging with limited image quality. The median angular luminosity distance and the typical DESI {\it r}-band FWHM of our sample is 43\,Mpc and 0.9\,arcsec, respectively. Thus the FWHM translates to a typical linear resolution of 0.2\,kpc. Considering the high quality of linear resolution of the DESI images and the typical bar size range of $0.5\text{\textendash}10$\,kpc found in nearby galaxies \citep{Erwin2005,Diaz-Garcia2016}, the bar structures are effectively spatially resolved in our nearby galaxy sample. We use the star-cleaned {\it r}-band images provided by \cite{Yu2023} for the our analysis, as the NIRCam of JWST will trace the rest-frame optical light at high redshifts. The star-cleaned images are essential for the creation of images of various resolution and S/N levels observed by JWST. These images are generated by replacing the flux emitted by sources other than the target galaxy with interpolated flux or flux in the rotational symmetric regions (details see \citealt{Yu2023}).

\begin{figure}[!h]
\begin{center}
\includegraphics[width=0.45\textwidth]{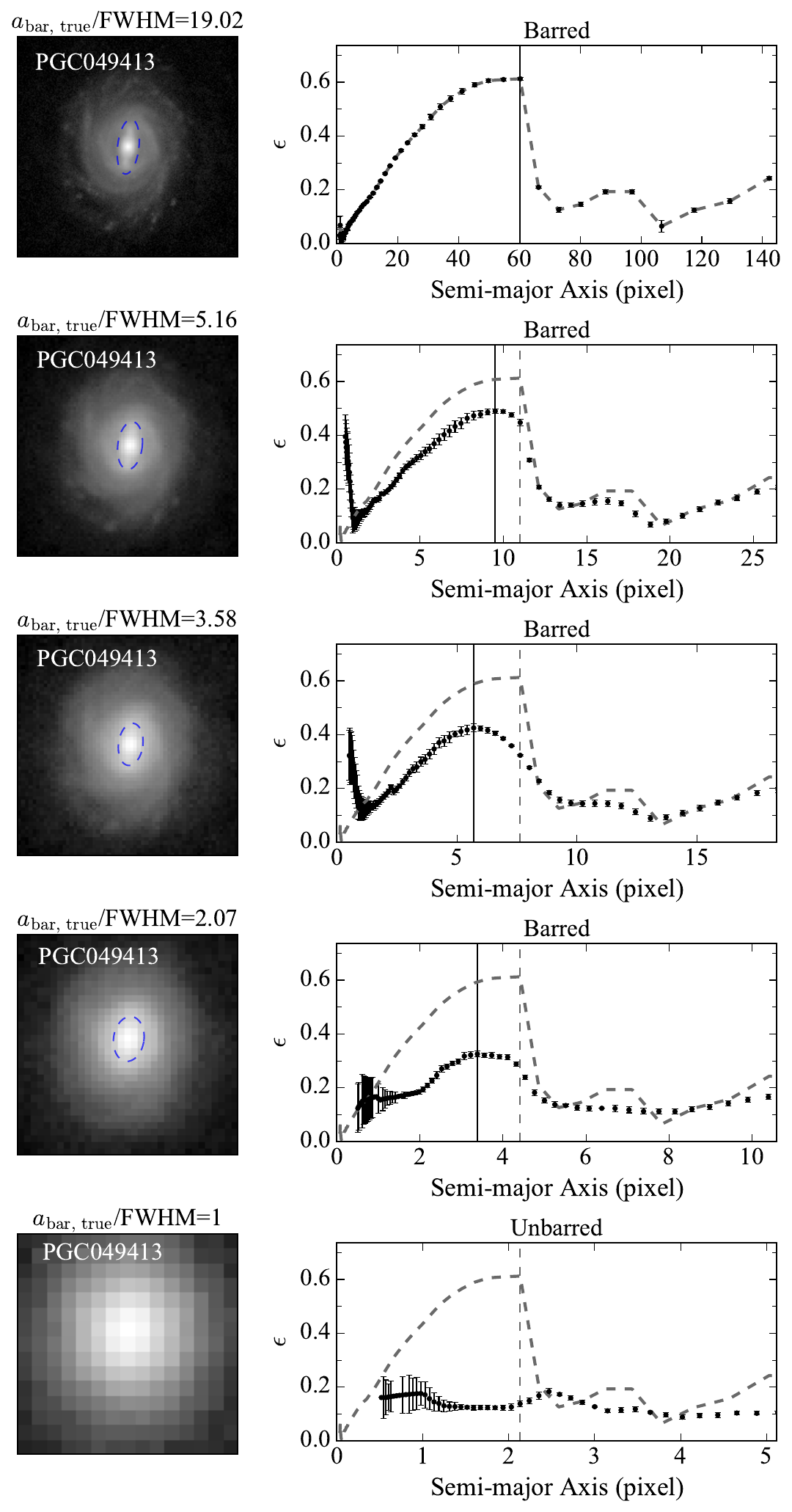}
\caption{Illustration of the impact of decreasing resolution on the bar analysis, using PGC049413 as an example. On the left are the galaxy images, with their respective derived $\epsilon$ profiles to the right. At the top of each image, the ratio of $a_{\rm{bar,\,true}}$ to FWHM is indicated. The first row shows the results from DESI, succeeded by the results from low-resolution F200W images in the rows below. The DESI $\epsilon$ profiles are plotted with grey dashed curves in the right panels, with the intrinsic bar position denoted as a grey vertical dashed line. When a bar is successfully identified, it is marked on the image by an ellipse and indicated on the profile by a vertical solid line. 
}
\label{fig:eprof_1}
\end{center}
\end{figure}

\begin{figure}[!h]
\begin{center}
\includegraphics[width=0.45\textwidth]{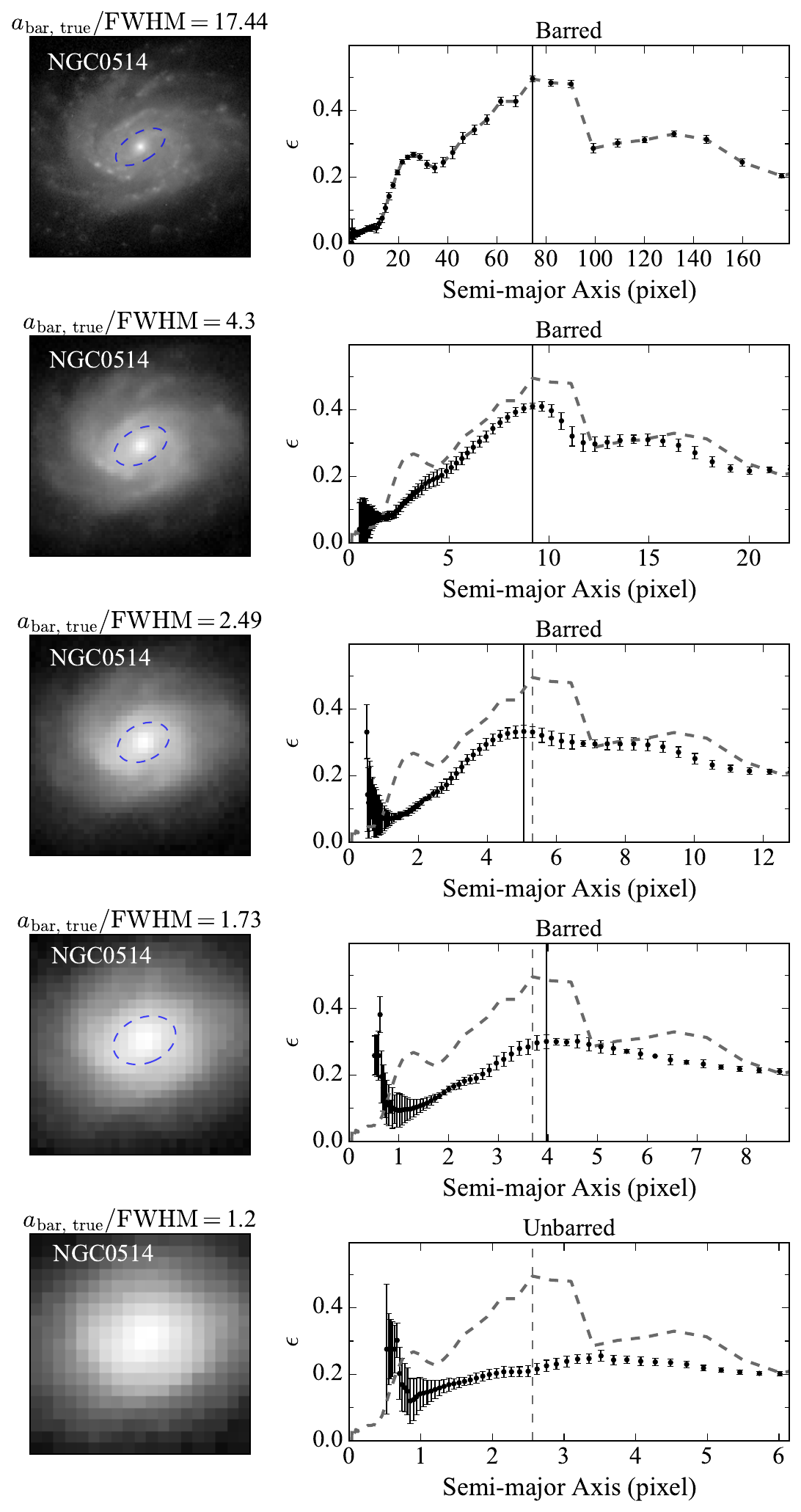}
\caption{Illustration of the impact of decreasing resolution on the bar analysis, using NGC0514 as another example. Details are the same as Fig.~\ref{fig:eprof_1}.
}
\label{fig:eprof_2}
\end{center}
\end{figure}

Several methods have been established to identify bars and measure their properties, including visual inspection \citep[e.g.,][]{vaucouleurs1991,NairandAbraham2010,Herrera2015}, ellipse fitting method \citep[e.g.,][]{Mendez2007,MarinovaandJogee2007,Sheth2008, YuHo2020}, Fourier Analysis \citep[e.g.,][]{Ohta1990,Elmegreen1985,Aguerri1998}, and two dimensional decomposition \citep[e.g.,][]{Gadotti2009,Laurikainen2005,Salo2015}. There are always limitations and potential uncertainties associated with these techniques, and no single method can guarantee perfect results \citep{AthanassoulaandMisitiotis2002,Aguerri2009,Lee2019}. We use ellipse fitting method to analyze bars. The isophote ellipticity ($\epsilon$) has been shown to increase with radius within regions dominated by bars, beyond which the $\epsilon$ would drop \citep[][]{Mendez2007, Aguerri2009}. However, in some barred galaxies, especially the SAB galaxies (which represent an intermediate class between spirals and those with strong bars), this drop in the $\epsilon$ profile can be absent, as the bar is not strictly straight, reducing the difference in $\epsilon$ between the bar and underlying disk. Another point to consider is the observational effect. For galaxies observed at high redshifts, the drop in the $\epsilon$ profile can be smoothed out by the PSF, as will be illustrated in Fig.~\ref{fig:eprof_1}, \ref{fig:eprof_2}, and \ref{fig:zhi_eprof_1}. Therefore, we adopt the strategy suggested by \cite{ErwinandSparke2003} to seek out peaks in the $\epsilon$ profile as a signature of bars or potential bar candidates. While this approach is effective, we opted not to account for possible twists in the position angle (PA) profile, but supplement our approach with visual inspections to ascertain the presence of bars.

We use the {\tt ellipse} task from {\tt photutils}\footnote{https://photutils.readthedocs.io} twice for each image to perfrom isophote fitting. First, we set the center, $\epsilon$, and PA as free parameters and then determine the galaxy center by averaging the centers of the resulting isophotes in the inner region. Second, we fix the center to derive the profiles of $\epsilon$ and PA. For each $\epsilon$ profile, we select the bar candidates by search for local peaks greater than 0.1 in the $\epsilon$ profile. For each bar candidate, we visually check if it represents a bar instead of dust lanes, star forming regions, or spiral arms. If one of the candidates is confirmed as a bar, the galaxy is classified as barred galaxies and the semi-major axis (SMA), $\epsilon$, and PA corresponding to the selected peak (denoted as $a_{\rm bar,\,true}$, $\epsilon_{\rm bar,\,true}$, and $\rm PA_{\rm bar,\,true}$, respectively) are taken as the measures of the bar properties. 
The subscript of ``true'' is used to distinguish these values from those obtained by analyzing simulated degraded images in Sect.~\ref{sect:sim}.
The $a_{\rm bar,\,true}$ describes the apparent bar size. For several galaxies where a bar is obvious by visual inspection, the $\epsilon$ profile doesn't exhibit a peak because the ellipticity of the bar is comparable to that of the disk. We manually determine the SMA that best characterizes the bar and subsequently estimate its properties. The remaining galaxies are classified as unbarred galaxies.

\begin{figure}
\begin{center}
\includegraphics[width=0.45\textwidth]{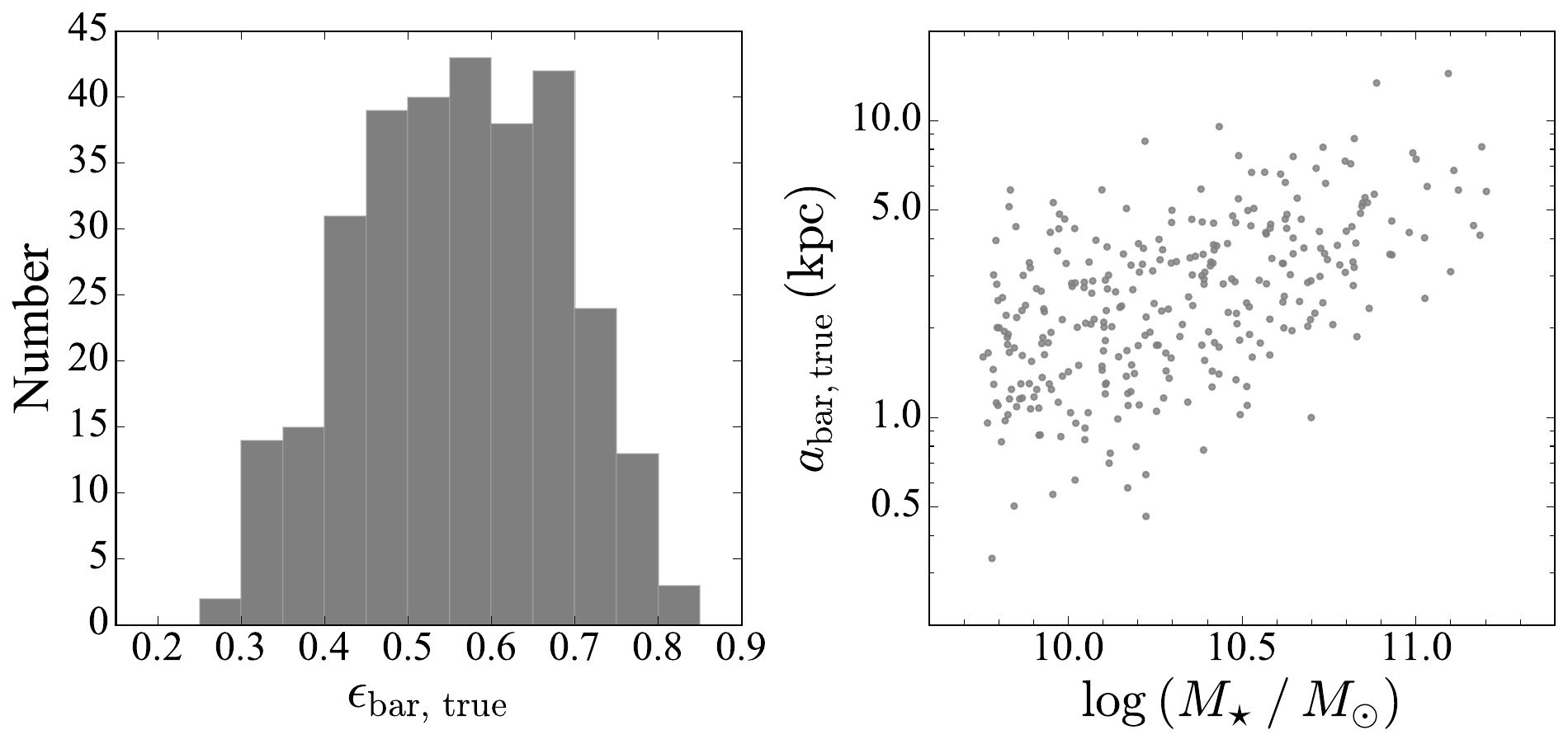}
\caption{Basic properties of the barred galaxies in our sample.
The left panel illustrates the distribution of ellipticity of bars ($\epsilon_{\rm bar,\,true}$), while the right panel shows the  projected bar size as semi-major axis ($a_{\rm bar,\,true}$) versus stellar mass ($M_{\star}$).}
\label{fig:summary_desi}
\end{center} 
\end{figure}

Out of 448 spiral galaxies, 304 are classified as barred, yielding a bar fraction of 68\% that, albeit being slightly lower, is consistent with previous studies identifying bars through visual inspection \citep[e.g.,][]{vaucouleurs1991, MarinovaandJogee2007, Buta2015}. This fraction is marginally higher than those found in some studies using ellipse fitting methods \citep[e.g.,][]{Barazza2008, Aguerri2009}, which have missed barred galaxies, especially the SAB types, that don't exhibit a sudden drop in their $\epsilon$ profiles.  This calculated bar fraction of 68\% is denoted as $f_{\rm bar,\,true}$. Figure \ref{fig:summary_desi} summaries the basic bar properties in our sample. The distribution of $\epsilon_{\rm bar,\,true}$ is shown in the left panel, and the relationship between  $a_{\rm bar,\,true}$ and galaxy stellar mass is displayed in the right panel. Our sample covers a wide range of bars, with $\epsilon_{\rm bar,\,true}$ values spanning from 0.25 to 0.85 and $a_{\rm bar,\,true}$ ranging from 0.3 to 10 kpc. We demonstrate that more massive galaxies have longer $a_{\rm bar,\,true}$, consistent with previous work \citep[e.g.][]{Diaz-Garcia2016, Erwin2019, Kim2021}, though our analysis does not involve the deprojection process for the bars. We refer to these properties derived from DESI images as the true bar properties due to the high quality of the DESI images.

\section{Image simulations} \label{sect:sim}

Two approaches are employed to investigate the cosmological redshift effects on the detection and measurement of bars at high redshift with JWST. The first approach, for considering observational effects, is to use the {\it r}-band star-cleaned image of nearby galaxies to generate and study simulated images of various resolution and $S/N$ (Sect.~\ref{sect:N FWHM} and Sect.~\ref{sect:SimulatedSNR}). In the second approach, we analyse the simulated CEERS images that were created by taking into account for both observational effects and galaxy evolutions (Sect.~\ref{sect:ceers}). We re-identify and re-measure the bars present in these simulations to illustrate how resolution, noise, and their combined effects influence the analysis of bars in galaxies at high redshifts.

\subsection{Simulated low-resolution images} \label{sect:N FWHM}

The detectability of the bar is often gauged by the ratio of $a_{\rm{bar,\,true}}$ to PSF FWHM, that is $n=a_{\rm bar,\,true}/{\rm FWHM}$, however, the quantitative impact of this ratio remains to be elucidated. To observe rest-frame optical wavelength, we assume using the F115W, F150W, and F200W filters for studying bars in real observations at redshifts of $z<1$, $1\leq z<2$, and $2\leq z<3$, respectively.  Although the JWST F115W images can better resolve bar structures owing to their narrower PSF, they may miss bars at high redshifts due to a shift toward shorter rest-frame wavelengths. Consequently, the redder filters are utilized to mitigate this effect.  We thus examine three JWST PSFs corresponding to the filters F115W, F150W, and F200W, which have FWHMs of 0.037, 0.049, and 0.064 arcsec, respectively\footnote{https://jwst-docs.stsci.edu/jwst-near-infrared-camera/nircam-performance/nircam-point-spread-functions}. A pixel scale of 0.03 arcsec/pixel, consistent with the CEERS data release \citep{Bagley2023}, is adopted. 

To generate low-resolution images for understanding the impact of PSF smoothing, we start by downsizing the star-cleaned images to match an exponentially increasing sequence of $n$ values: $1.0, 1.2, 1.44, 1.73, 2.07, 2.49, 2.99, 3.58, 4.3, 5.16,$ and $10$. Correspondingly, the DESI PSFs are resized using the same scaling factor. Next, we generate JWST PSFs using {\tt WebbPSF} \citep{Perrin2014} and derived a kernel via Fourier transformation to transform the resized PSF to the JWST PSF.  Lastly, we convolve the downsized galaxy image with the kernel to obtain the simulated low-resolution image. These simulated images haves a very high $S/N$, allowing us to focus on examining the impacts of the PSF. We analyse the bar structure in each low-resolution image using the same ellipse fitting method outlined in Sect.~\ref{sect:obs}. The derived bar size, ellipticity, and position angle are donated as $a_{\rm obs}^n$, $\epsilon_{\rm obs}^n$, and ${\rm PA}_{\rm obs}^n$.

Figure~\ref{fig:eprof_1} uses PGC049413 as an example to illustrate the impact of decreasing resolution on the bar analysis, when the F200W PSF is considered. The images are shown in the left column, while the $\epsilon$ profiles are shown on the right. The top row shows the result derived from the DESI image. As the DESI image has quite high resolution, indicated by $a_{\rm bar,\,true} / \rm FWHM=19.02$, there is clearly a bar in the image and a peak or equivalently a drop in the $\epsilon$ profile. The peak or the drop, marked by the solid line in the profile, is selected to measure the bar. An ellipse with the resulting parameters is plotted on the image to illustrate the measurement. The subsequent rows present the results obtained from the simulated low-resolution images. To facilitate the comparison between the results before and after resolution degradation, we adjust the SMA of the DESI $\epsilon$ profile to match those from the low-resolution images, and then plot the adjusted profiles as grey dashed curves and intrinsic bar size as the vertical dashed line. As the resolution decreases to $a_{\rm bar,\,true} / \rm FWHM=5.16$ and 3.58, the bar structures in the images remain clearly visible despite the increasing blur. The persistent peak in the $\epsilon$ profile underscores the presence of the bar, though its amplitude decreases. This reduced peak amplitude indicates an underestimation of bar ellipticity. Meanwhile, the peak shifts inward, leading to an underestimation of measured bar size. Moreover, the sudden drop seen in the DESI $\epsilon$ profile softens at $a_{\rm bar,\,true} / \rm FWHM=5.16$ and is completely absent at $a_{\rm bar,\,true} / \rm FWHM=3.58$. This behavior leads us to use peak in $\epsilon$ profiles to identify bars, as described in Sect.~\ref{sect:obs}. These changes in the image and the $\epsilon$ profiles are caused by the PSF convolution, which rounds the bar structure and reduces its clarity at the edges.  When the resolution decreases to $a_{\rm bar,\,true} / \rm FWHM=2.07$, the image becomes more unclear, causing subtle structural details like spiral arms to significantly fade. However, the bar structure remains discernible in the image with a distinct peak in the $\epsilon$ profile. When the resolution further decreases to $a_{\rm bar,\,true} / \rm FWHM=1$, the image has become so blurred that any structure is completely invisible and the characteristic features corresponding to a bar in the $\epsilon$ profile disappears.

For simulated images of PGC049413, we note the significant decrease in the measured bar size. At $a_{\rm bar,\,true} / \text{FWHM} = 3.58$, the fractional difference between measured and intrinsic bar size yields $-25\%$. However, such cases are not typical. On average, lower resolution only reduces the measured bar size by a few percent when $a_{\text{bar}}/\text{FWHM} > 2$, as will be discussed in detail in Sect.~\ref{sect:results}.  Figure~\ref{fig:eprof_2} uses NGC0514, a SAB galaxy, as an example to illustrate the typical resolution effect. At $a_{\rm bar,\,true}/ \rm FWHM =4.3$, the bar size still matches the original size, while at $a_{\rm bar,\,true}/ \rm FWHM = 2.49$, the observed bar size is reduced by 5\%. However, at $a_{\rm bar,\,true}/ \rm FWHM =1.73$, the bar size is overestimated due to the image being blurred, causing part of the spiral structure to be recognized as the bar. At $a_{\rm bar,\,true}/ \rm FWHM = 1.2$, the bar structure is completely lost. Additionally, similar to the previous illustration of PGC049413, the amplitude of the $\epsilon$ profile gradually decreases as the resolution worsens.

\subsection{Simulated low-S$/$N images}\label{sect:SimulatedSNR}
To explore the influence of noise on the bar analysis under realistic observational conditions in CEERS, we first estimate the $S/N$ range of the galaxy images observed in the survey. We use the science data, error map, and source mask from the CEERS Data Release Version 0.6 (data reduction sees \citealt{Bagley2023})\footnote{https://ceers.github.io/dr06.html}. We use the catalog of \cite{Stefanon2017} to select galaxies with stellar mass $M_{\star} \geq 10^{9.75} M_{\odot}$ at redshifts of $0.75 \leq z \leq 3.0$, and then use {\tt sep} \citep{Bertin1996, Barbary2016} to generate a mask for each galaxy. Images of galaxies that are severely contaminated by close sources are excluded. We calculate the sky uncertainty through {\tt Autoprof} \citep{Stone2021} and derive the galaxy flux uncertainty from the error map. We calculate the map of $S/N$ for each galaxy by dividing the galaxy flux by the flux uncertainty pixel by pixel, and compute average $S/N$ over an elliptical aperture with SMA of galaxy Petrosian radius, and with galaxy $\epsilon$ and PA. This average $S/N$ value is considered as the $S/N$ of this galaxy. The rationale for calculating the $S/N$ averaged over the pixels occupied by the galaxy, instead of simply dividing the total galaxy flux by its uncertainty (which treats the galaxy like a point source), is that the $S/N$ of individual pixels can vary throughout the galaxy. Since bars are extended structures, averaging the $S/N$ over the pixels that encompass the bars provides a more accurate representation of the signal strength of the bars relative to the noise.

As a result, for CEERS galaxies observed in the F115W ($0.75<z<1$), F150W ($1<z<2$), and F200W ($2<z<3$) filters, the median $S/N$ values are 20.0, 11.4, and 10.0, respectively. The corresponding stardard deviations are 11.5, 9.5, 9.0. Moreover, over 95\% of galaxies in each filter exhibit a $S/N$ greater 3. Therefore, to investigate whether and how the typical noise level in CEERS influences our bar analysis, we use the high-$S/N$ simulated image of $a_{\rm bar,\,true}/{\rm FWHM}=4.3$ to generate, for each galaxy, simulated images with an exponentially increasing $S/N$: 3, 4.2, 5.9, 8.2, 11.5, 16.1, 22.6, 31.6, 44.3, and 62.0. The ratio $a_{\rm bar,\,true}/{\rm FWHM}$ is fixed for isolating the noise effect. We have verified that using images of other $a_{\rm bar,\,true}/{\rm FWHM}$ to examine noise effects does not significantly affect our results. To make the image noisy, we rescale the image flux with a scaling factor, derive flux noise, and add the flux noise as well as a background map to the flux-rescaled image. In this process, we use the median ratio of galaxy flux to flux variance and the patch of cleaned real CEERS background, as provided by \cite{Yu2023}, to compute the galaxy flux noise and to represent the background map, respectively. The scaling factor is iteratively adjusted to match our desired $S/N$. We apply the same ellipse fitting method as described in Sect.~\ref{sect:obs} to identify and quantify bars. The derived bar size, ellipticity, and position angle from these images of various $S/N$ are denoted as $a_{\rm obs}^{S/N}$, $\epsilon_{\rm obs}^{S/N}$, and ${\rm PA}_{\rm obs}^{S/N}$.

\begin{figure}
\begin{center}
\includegraphics[width=0.45\textwidth]{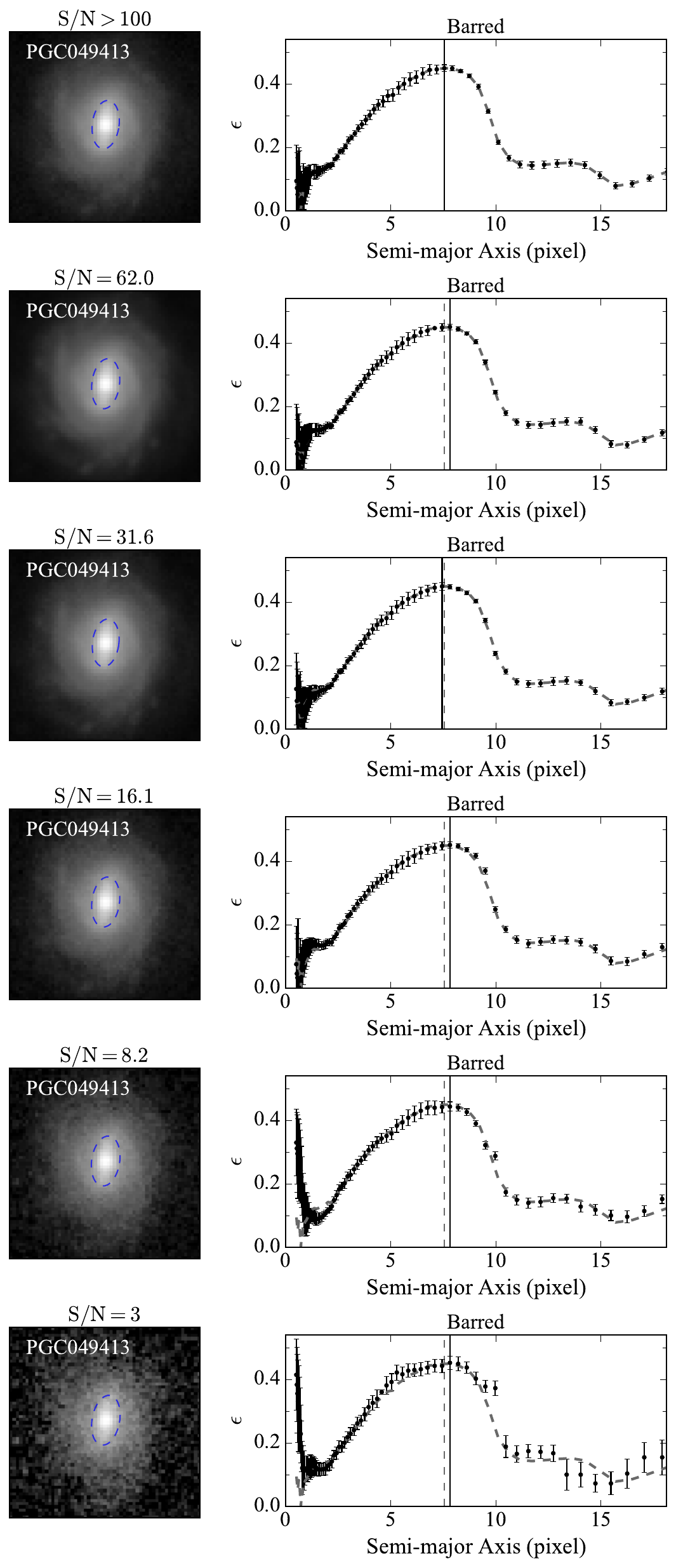}
\caption{Illustration of the impact of noise on the bar analysis. The symbols are consistent with Fig. \ref{fig:eprof_1}. The top row now displays the noiseless simulated image of $a_{\rm bar,\,true}/{\rm FWHM}=4.3$ and its profile, while the subsequent rows show the simulated low-$S/N$ images and their profiles.} 
\label{fig:SNR_eprof_1}
\end{center}
\end{figure}

With the layout consistent with Fig.~\ref{fig:eprof_1}, Fig.~\ref{fig:SNR_eprof_1} uses PGC049413 as an example to illustrate the impact of noise. When $S/N\geq 16.1$ (top four rows), noise hardly affects the images, keeping the bar structure clear and the corresponding $\epsilon$ profiles unchanged. As the S$/$N drops to 8.2 and further to 3 (bottom two rows), the influence of noise becomes a bit more significant, resulting in a decrease in image clarity and greater uncertainty in the $\epsilon$ profile. Despite this, the bar structure is still discernible, and the general shape of the $\epsilon$ profiles undergoes only minor change.  Measurements of the bar size and ellipticity from images with low $S/N$ are consistent with those from images with high $S/N$.  Therefore, the bar identification and quantification for PGC049413 are not significantly affected by the noise within the typical $S/N$ range observed in the CEERS field.

\begin{figure}
\begin{center}
\includegraphics[width=0.45\textwidth]{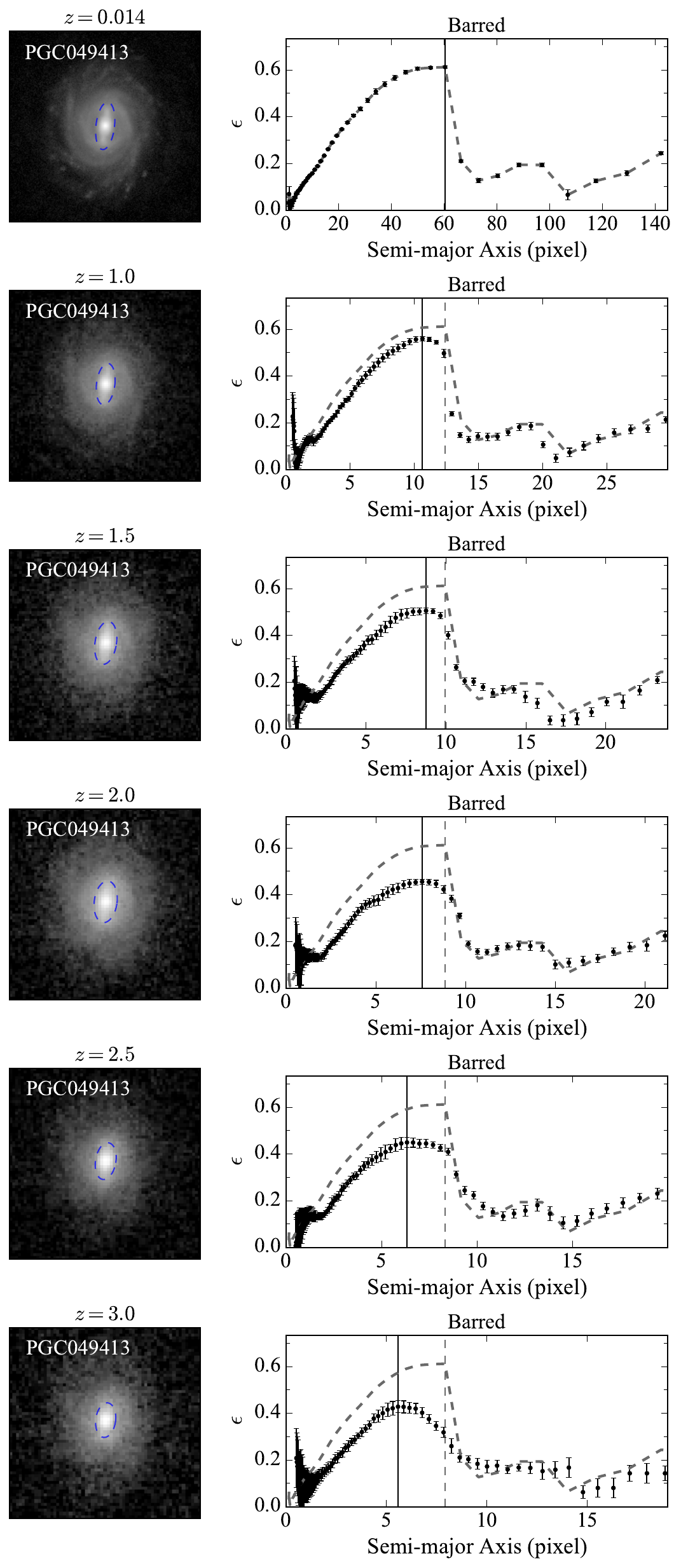}
\caption{Illustration of the impact of redshift effects on the bar analysis. The redshift effects include observational effects and galaxy (and bar) evolution. The symbols are consistent with Fig. \ref{fig:eprof_1}, except that simulated CEERS images at various redshifts and their results are shown.} 
\label{fig:zhi_eprof_1}
\end{center}
\end{figure}

\subsection{Simulated galaxy images observed in CEERS}\label{sect:ceers}

To examine the cosmological redshift effect, that is a combination of resolution and noise effects in a specific way, simulated CEERS images with both observational effects and evolution effects considered are essential. \cite{Yu2023} used multi-waveband images of a sample of nearby DESI galaxies to generate artificially redshifted images observed in JWST CEERS at $z=0.75$, 1.0, 1.25, 1.5, 1.75, 2.0, 2.25, 2.5, 2.75, and 3.0. The F115W filter is used for $z = 0.75$ and 1, the F150W filter is used for $z = 1.25$, 1.5, and 1.75, and the F200W filter is used for $z = 2.0$ to 3.0. Their image simulation procedure involves spectral change, cosmological surface brightness dimming, luminosity evolution, physical disk size evolution, shrinking in angular size due to distance, decrease in resolution, and increase in noise level (for details, see \citealt{Yu2023}). In particular, PSFs are matched to those generated using {\tt WebbPSF} and a cleaned blank CEERS background is employed. A bar evolution model is incorporated. With increasing redshift, the physical size of bars becomes shorter as the disk size becomes smaller following the galaxy size evolution derived by \cite{Vanderwel2014}.  Applying a galaxy evolution model is a separate, additional consideration, apart from the basic resolution and $S/N$ effects. This dataset fits our scientific goal of understanding the redshift effects on the bar measurements. We analyze each simulated CEERS image using the ellipse fitting method described in Sect.~\ref{sect:obs}. The resultant bar size, ellipticity, and position angle are denoted as $a^{z}_{\rm obs}$, $\epsilon^{z}_{\rm obs}$, and ${\rm PA}^{z}_{\rm obs}$.

Figure \ref{fig:zhi_eprof_1} showcases the redshift effects on bar measurement. The top row displays results derived from the DESI image, while the subsequent rows present those from the simulated CEERS images at high redshifts. The galaxy structures in the DESI image are clear and the drop or peak in the $\epsilon$ profile is prominent. When the galaxy is artificially moved to $z=1$ and then further to $z=3$, the galaxy image becomes increasingly blurred and noisy. Meanwhile, the $\epsilon$ profiles tend to flatten, with the sudden drop in the $\epsilon$ profile disappearing entirely at $z\geq 1.5$. The flattening in the $\epsilon$ profiles is due to the decreasing $a_{\rm bar,\,true}/{\rm FHWM}$ with increasing redshifts, consistent with the expectation based on the resolution effect discussed in Sect.~\ref{sect:N FWHM}. As anticipated from Sect.~\ref{sect:SimulatedSNR}, the noise only make the profiles at high redshifts slightly more chaotic than low-redshift results. Together with the results shown in Sections \ref{sect:N FWHM} and \ref{sect:SimulatedSNR}, Figure~\ref{fig:zhi_eprof_1} suggests that the noise has a minimal impact on the analysis of bars observed in the CEERS field for the mass and redshift ranges we considered, and the resolution effect is the predominant factor. While the peak corresponding to the bar is reduced gradually with increasing redshift, the bar remains detectable both in the image and in the $\epsilon$ profile for PGC049413. Nevertheless, it's evident that the difference between bar size and bar ellipticity measured from the simulated high-redshift images and their true values are increasing.

\begin{figure}
\begin{center}
\includegraphics[width=0.45\textwidth]{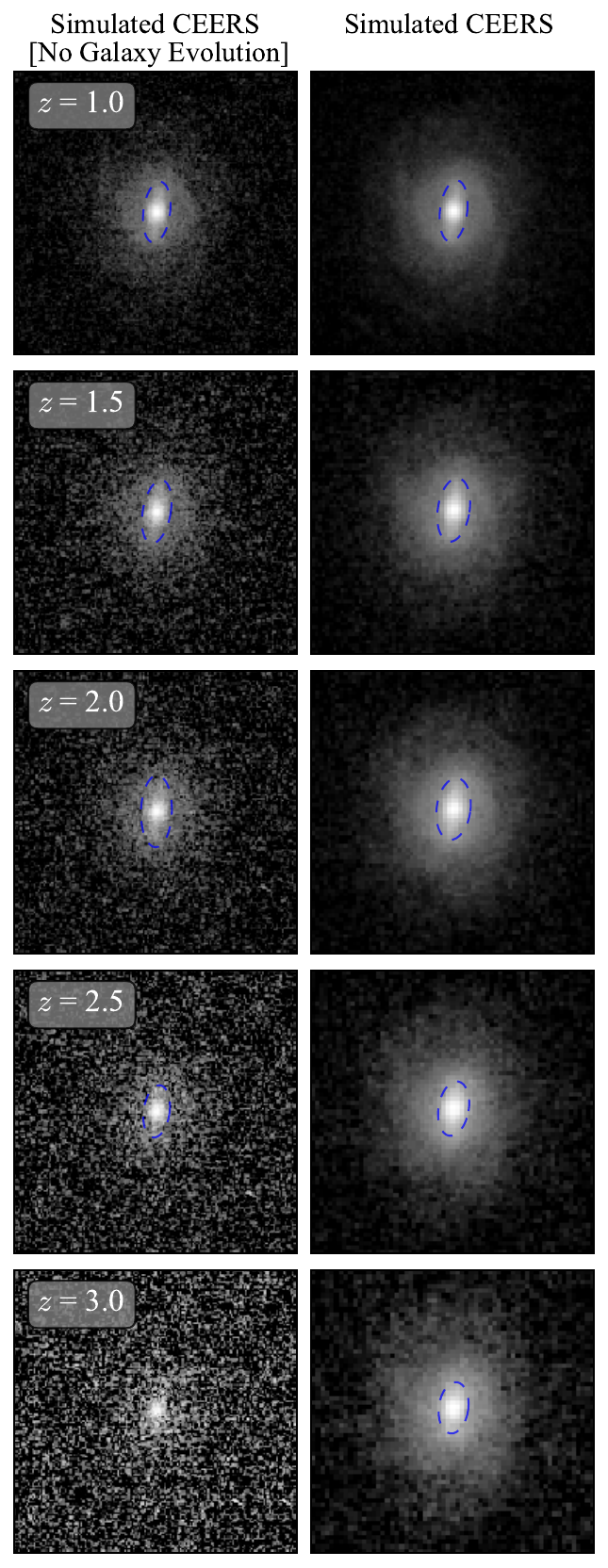}
\caption{Illustration of the simulated CEERS image with the adoption of a galaxy evolution model (on the right) as well as the one without (on the left) using PGC049413 as an example, presented as the redshift increases. The blue ellipse marks the detected bar structure in the image if a bar is identified.}
\label{fig:Comp_img}
\end{center}
\end{figure}

To illustrate the necessity of incorporating galaxy evolutions in generating simulated CEERS images, we created another set of simulated images considering only observational effects without any evolutions. Figure~\ref{fig:Comp_img} compares the two sets of simulated data using PGC049413 as an example. The left column shows the simulated images without any galaxy evolution models, while the right column shows the simulated images with the galaxy evolution models. As $z$ increases, the bar structure gradually fades due to noise. At $z=3.0$, the bar is still visible from the simulated CEERS image with evolutions but appears completely faded into the background in images without evolutions. The $S/N$ of the images without evolution models (in intrinsic surface brightness) is significantly reduced due to the cosmological dimming effect on observed galaxy surface brightness, which become fainter by $2.5\log(1+z)^{3}$ at high redshifts (AB magnitude is used). However, the intrinsic surface brightness has been found to brighten with increasing redshift \citep[e.g.,][]{Barden2005, Sobral2013}, which is caused by a combination of luminosity evolution and size evolution \citep{Yu2023}. 
We calculated the $S/N$ for these images and found that, from $z=1.25$, the $S/N$ falls below 3 and is significantly lower than that for real CEERS images. At $z=3.0$, the median value of $S/N$ yields only 0.54. The low $S/N$ causes a large number of bars to be undetected. These noisy images are therefore inconsistent with the real CEERS galaxy images presenting quite good $S/N$ to resolve galaxy structures, validating our procedure to generate simulated CEERS images by incorporating galaxy evolution models.

We give a caution that our bar evolution model may not be correct. Our input bar model assume the bar size related to disk size remains unchanged across redshift. Nevertheless, such a ratio changes by a factor of $\sim$2 from $z\sim3$ to 0 in simulations \citep{Anderson2024}. Future studies comparing fraction and size of bars observed in JWST images with simulated images may give a better constraint on the bar evolution model.

\begin{figure}
\begin{center}
\centering
\includegraphics[width=0.45\textwidth]{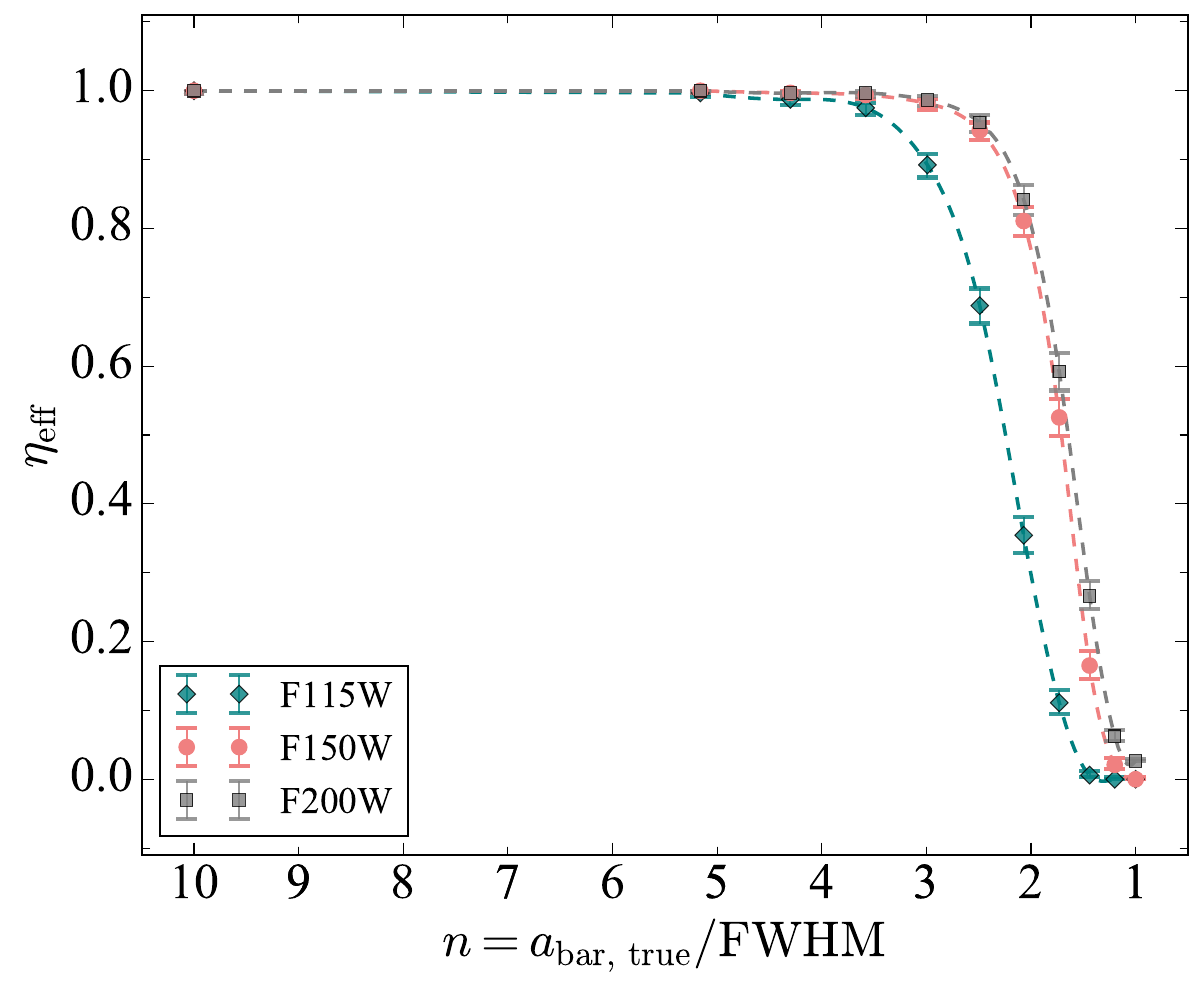}
\caption{Effectiveness of the method of detecting bars, $\eta_{\rm eff}$, as a function of resolution $n=a_{\rm bar,\,true}/{\rm FWHM}$. The green diamonds, red circles, and grey squares represent the results based on the F115W, F150W, and F200W filters, respectively. } 
\label{fig:BarfracN}
\end{center}
\end{figure}

\begin{figure*}
\begin{center}
\centering
\includegraphics[width=1\textwidth]{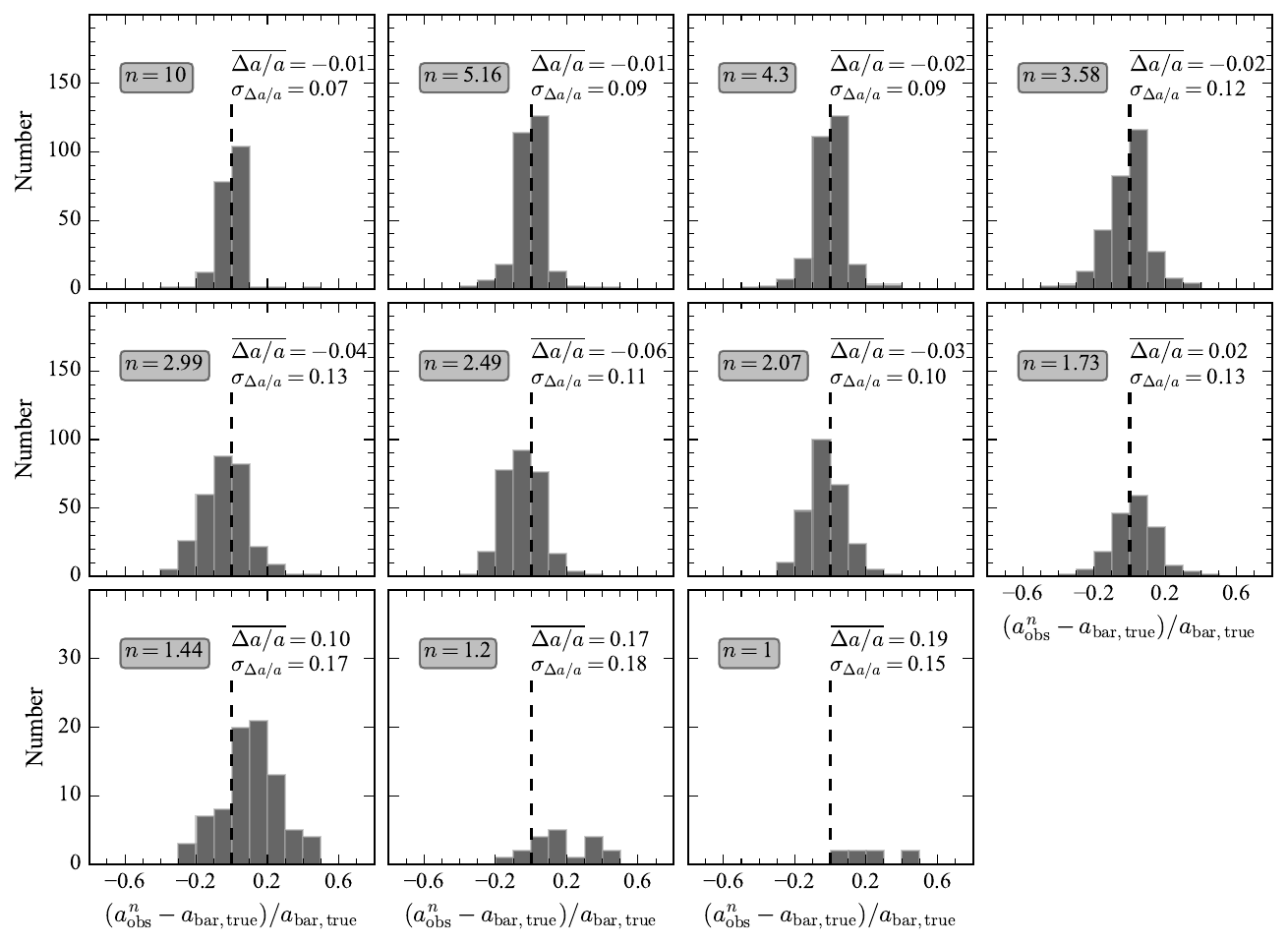}
\caption{Distribution of the fractional difference ${\Delta a^n_{\rm obs} / a_{\rm bar,\,true}}$, where $\Delta a^n_{\rm obs} = (a_{\rm obs}^n - a_{\rm bar,\,true})$, at various resolution levels $n=a_{\rm bar,\,true}/{\rm FWHM}$. The $a_{\rm bar,\,true}$ denotes the intrinsic value. Results for the F200W PSF are present. The averaged fractional difference ($\overline{{\Delta a/a}}$) and standard deviation ($\sigma_{\Delta a/a}$) are indicated at the top of each panel. The vertical black dashed line in each panel marks $a^n_{\rm obs}=a_{\rm bar,\,true}$.}
\label{fig:RwithN}
\end{center}
\end{figure*}

\begin{figure*}
\begin{center}
\centering
\raggedright
\includegraphics[width=1\textwidth]{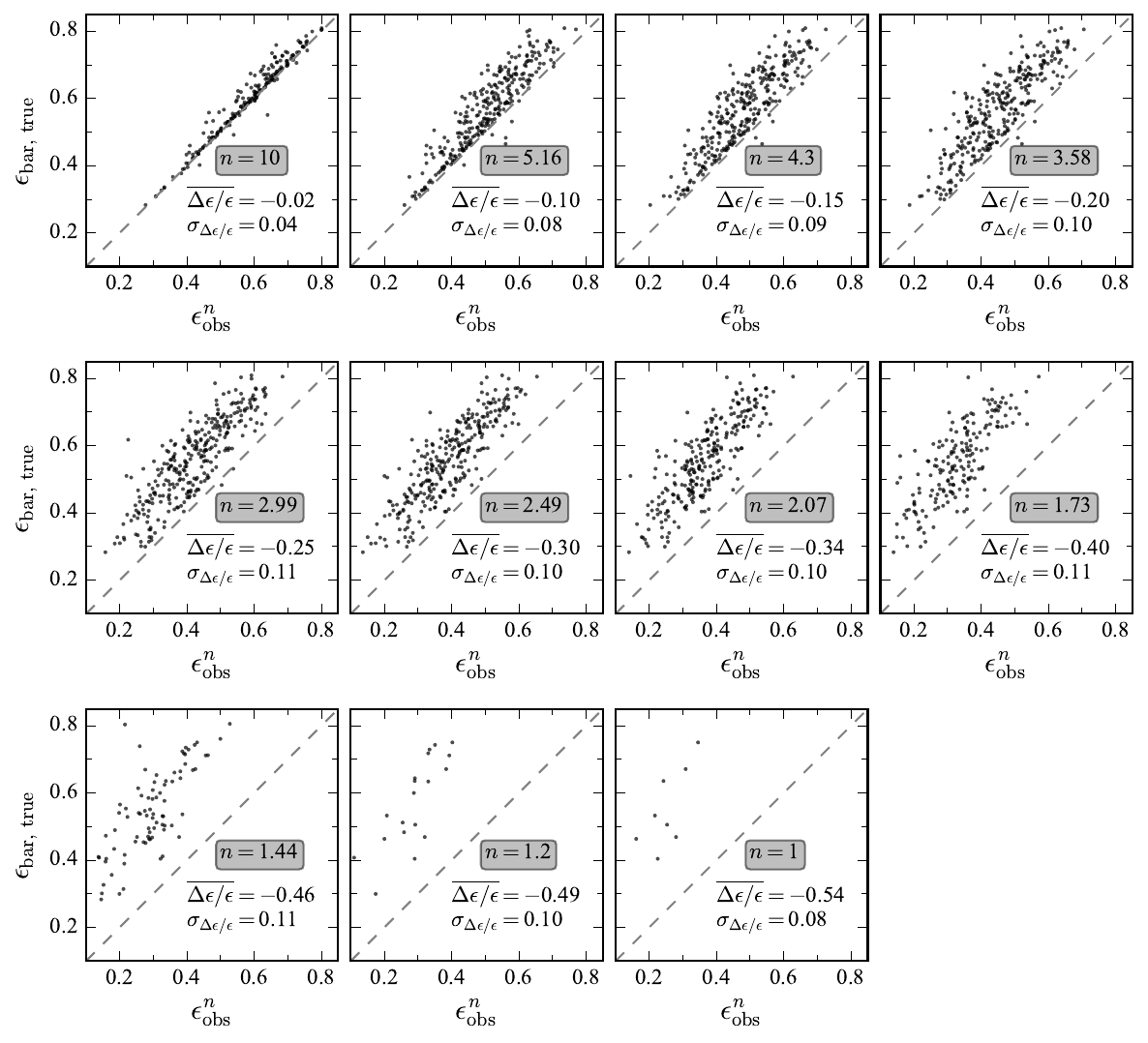}
\caption{Comparison between $\epsilon_{\rm obs}^{n}$ and $\epsilon_{\rm bar,\,true}$ at each resolution level $n=a_{\rm bar,\,true}/{\rm FWHM}$. Results for the F200W filter are present. The fractional difference between $\epsilon_{\rm obs}^{n}$ and $\epsilon_{\rm bar,\,true}$ is denoted as $\Delta{\epsilon}/\epsilon = (\epsilon_{\rm obs}^n-\epsilon_{\rm bar, True})/\epsilon_{\rm bar, True}$. The mean value ($\overline{{\Delta \epsilon/\epsilon}}$) and standard deviation ($\sigma_{\Delta \epsilon/\epsilon}$) of the fractional difference are indicated in each panel. 
The one-to-one relation is displayed by a dashed line.} 
\label{fig:ewithN}
\end{center}
\end{figure*}

\begin{figure*}
\begin{center}
\centering

\includegraphics[width=1\textwidth]{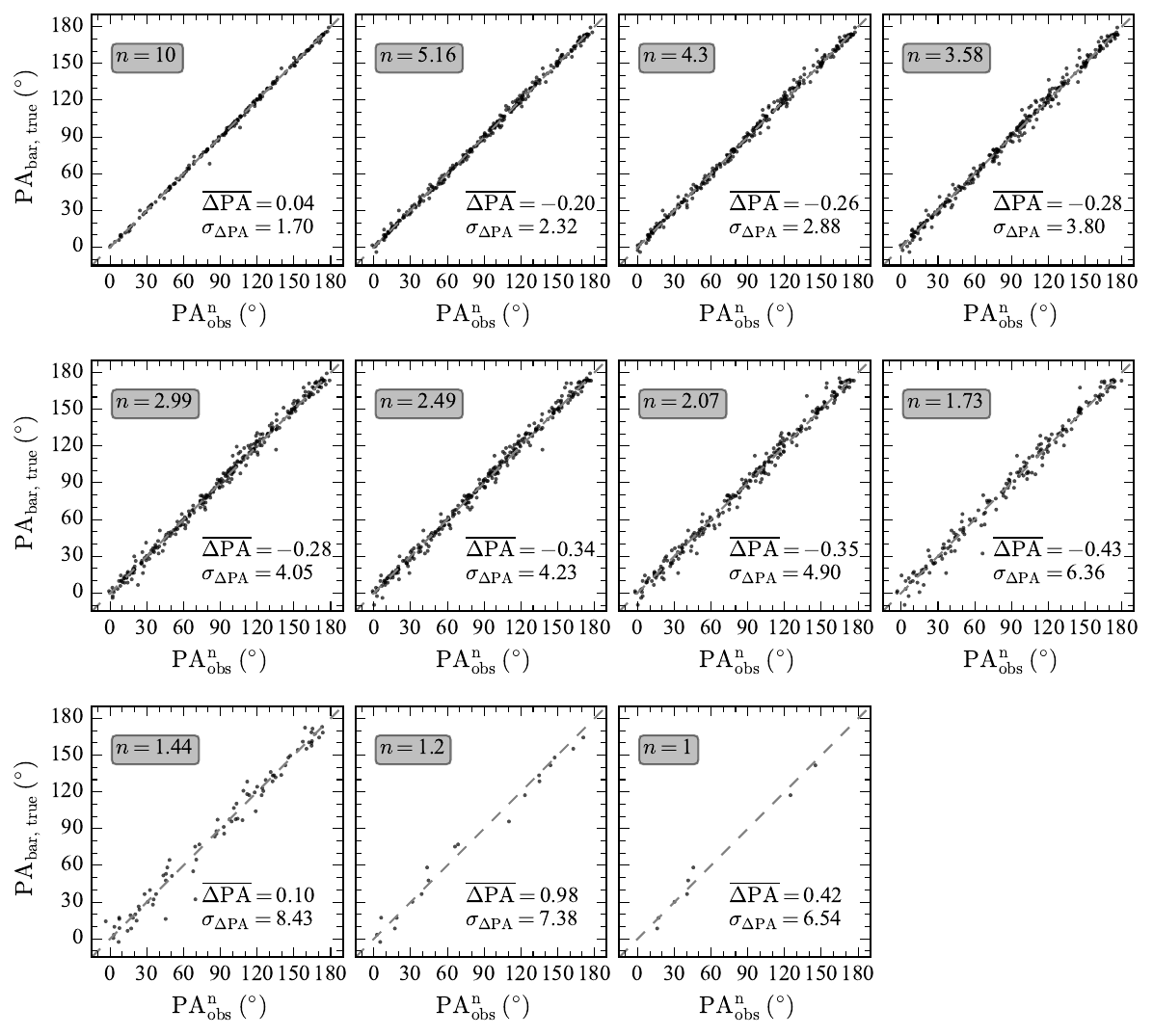}
\caption{
Comparison between ${\rm PA}^n_{\rm obs}$ and ${\rm PA}_{\rm bar,\,true}$ at each resolution level $n=a_{\rm bar,\,true}/{\rm FWHM}$. Results for the F200W filter are present. The difference between ${\rm PA}^n_{\rm obs}$ and ${\rm PA}_{\rm bar,\,true}$ is denoted as $\Delta{{\rm PA}} = {\rm PA}^n_{\rm obs}-{\rm PA}_{\rm bar,\,true}$. The mean difference ($\overline{{\Delta {\rm PA}}}$) and standard deviation ($\sigma_{\Delta {\rm PA}}$) are indicated in each panel. 
The one-to-one relation is displayed by a dashed line.
} 
\label{fig:pawithN}
\end{center}
\end{figure*}

\section{Measurement robustness of bar structures}\label{sect:results}

As illustrated in Sect.~\ref{sect:sim} through a representative example, the resolution limitation can significantly influence the identification and quantification of bars observed in the JWST CEERS field, while the effects of noise are minimal. These factors introduce considerable impact for galaxies at high redshifts. In this Section, we quantify these effects using the full sample of 448 galaxies.

\subsection{The effect of resolution}\label{results of resolution}

As resolution decreases, bars may not be detected. For each set of simulated low-resolution images with the specific value of $a_{\rm bar,\,true}/{\rm FWHM}$,   we determine the method effectiveness of detecting bars ($\eta_{\rm eff}$), defined as:
\begin{equation}\label{eff}
  \eta_{\rm eff}=\frac{\rm Number~of~detected~bars}{\rm All~bars}.
\end{equation}
\noindent
We plot the $\eta_{\rm eff}$ as a function of resolution ($n=a_{\rm bar,\,true}/{\rm FWHM}$) in Fig.~\ref{fig:BarfracN}, where  the green diamonds, red circles, and grey squares denoting the results for F115W, F150W, and F200W filters, respectively. We calculate the error of each $\eta_{\rm eff}$ (and bar fraction) using the  Wilson interval. With the $a_{\rm bar,\,true}/{\rm FWHM}$ decreasing from 10 to 1, the $\eta_{\rm eff}$ declines from 100\% to nearly 0. However, the profile shape varies across different bands. For the F115W band, the $\eta_{\rm eff}$ stays at $\sim$\,100\% until reaching $a_{\rm bar,\,true}/{\rm FWHM}\approx3$, after which it experiences a sharp decline.  We calculate the $a_{\rm bar,\,true}/{\rm FWHM}$ corresponding to  effectiveness of detecting bars of 50\% through interpolation and obtain 2.47.  For F150W or F200W band, the $\eta_{\rm eff}$ does not shown the sharp decline until reaching $a_{\rm bar,\,true}/{\rm FWHM}\approx2$. For F150W and F200W band, the 50\% effectiveness of detecting bars corresponds to $a_{\rm bar,\,true}/{\rm FWHM}=1.71$ and 1.65, respectively. Compared to F150W and F200W bands, the F115W-band $\eta_{\rm eff}$ trend has the sharp drop at higher $a_{\rm bar,\,true}/{\rm FWHM}$. This is because, given the pixel size of 0.03 arcsec/pixel, the F115W PSF FWHM gives $\sim$\,1.2 pixels, significantly below the 2-pixel threshold required for Nyquist sampling. This inadequacy in sampling causes the F115W image to lose small-scale structure information, leading some short basr to become undetectable.  Our results suggest that for images with a Nyquist-sampled PSF (such as JWST PSFs in the F200W, F277W, F356W, and F444W band), the critical bar size for detecting bars is $2\times {\rm FWHM}$, providing a quantitative justification for the empirical choice of $a_{\rm bar,\,true}=2 \times {\rm FWHM}$ as the bar size threshold for bar detection by \cite{Erwin2018}. \cite{Erwin2018} determine the factor of 2 based on the fact that almost all the bars in the S$^4$G sample are larger than $2 \times {\rm FWHM}$.

\begin{figure}
\begin{center}
\centering
\includegraphics[width=0.45\textwidth]{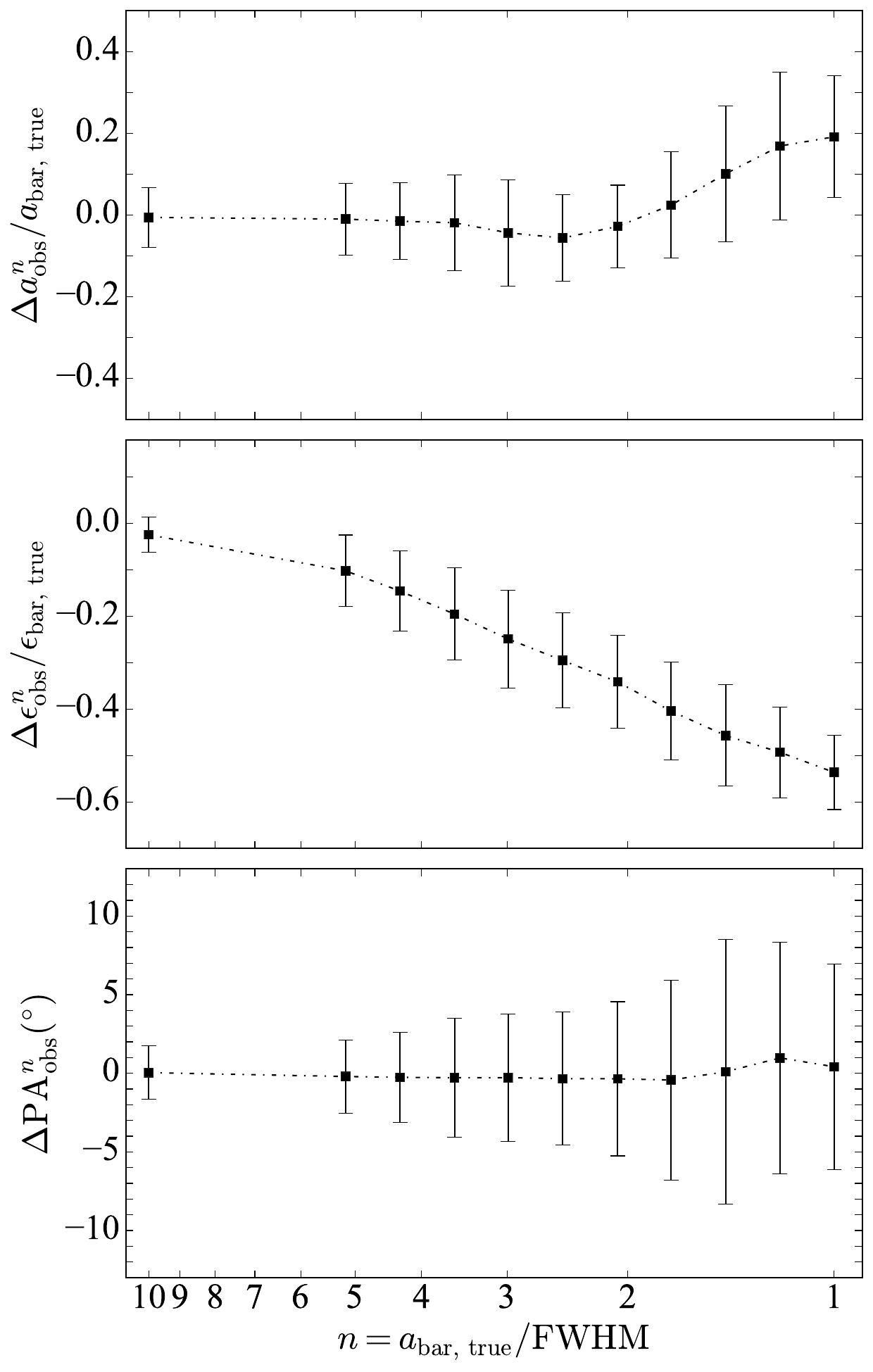}
\caption{
Change or fractional change between measure parameters and their intrinsic values as a function of resolution $n=a_{\rm bar,\,true}/{\rm FWHM}$. Results for the F200W filter are present.} 
\label{fig:delvsN}
\end{center}
\end{figure}

To understand how resolution can quantitatively impact the measured bar properties, we compare the properties ($a_{\rm obs}^n$, $\epsilon_{\rm obs}^{n}$, and ${\rm PA}_{\rm obs}^{n}$) derived from the low-resolution images with the intrinsic bar properties ($a_{\rm bar,\,true}$, $\epsilon_{\rm bar,\,true}$, and ${\rm PA}_{\rm bar,\,true}$) obtained from DESI images. We use the fractional difference ${{\Delta} a_{\rm obs}^{n} / a_{\rm bar,\,true}}$, where ${\Delta} a_{\rm obs}^{n} = (a_{\rm obs}^{n} - a_{\rm bar,\,true})$, to quantify the deviation in measured bar size as the resolution, $n=a_{\rm bar,\,true}/{\rm FWHM}$, decreases. The number distributions of ${\Delta a_{\rm obs}^{n} / a_{\rm bar,\,true}}$ at different $n$ are shown in Fig.~\ref{fig:RwithN}. Results for the F200W band are shown. In each panel, the $n$, the mean $\Delta a_{\rm obs}^{n}$, denoted as $\overline{{{\Delta}a/a}}$, and the standard deviation of $\Delta a_{\rm obs}^{n}$, denoted as ${\sigma}_{{\Delta}a/a}$, are present. The vertical dashed line marks the location of $a_{\rm obs}^n = a_{\rm bar,\,true}$. The overall distribution of ${{\Delta}a_{\rm obs}^n/a_{\rm bar,\,true}}$ is approximately symmetric. As the $n$ decreases from 10 to 2.49, the measured bar size tends to be more and more underestimated, though slightly, by from 1\% to 6\%. Such an underestimation is caused by PSF smoothing as demonstrated in Fig.~\ref{fig:eprof_1}.  However, as $n$ decreases from 1.73 to 1, the measured bar size tends to be more and more overestimated with lower resolution, by from 2\% to 19\%.  This can be clarified by the fact that, under relatively poor resolution conditions, various structures such as bars, spiral arms, and disks tend to become mixed. As a result, it becomes challenging to distinguish the bar distinctly, potentially leading to an overestimation of the measured bar size. The ${\Delta a_{\rm obs}^{n} / a_{\rm bar,\,true}}$ as a function of $n$ can be found in Fig.~\ref{fig:delvsN}.

In Fig. \ref{fig:ewithN}, we compare the intrinsic bar ellipcity $\epsilon_{\rm bar,\,true}$ with the measure bar ellipcity $\epsilon_{\rm obs}^{n}$ that are obtained from images at each resolution level $n=a_{\rm bar,\,true}/{\rm FWHM}$. The fractional difference is calculated as $\Delta{\epsilon}/\epsilon = (\epsilon_{\rm obs}^{n}-\epsilon_{\rm bar,\,true})/\epsilon_{\rm bar,\,true}$. Their mean value ($\overline{{\Delta \epsilon/\epsilon}}$) and standard deviation ($\sigma_{\Delta \epsilon/\epsilon}$) is indicated in each panel. As the resolution decreases from $n=10$ to 1, there is a clear trend that the data point distribution gradually shifts towards lower $\epsilon_{\rm obs}^{n}$. This trend can be quantitatively observed by examining the $\overline{\Delta\epsilon/\epsilon}$, which decreases from $-2\%$ to $-54\%$. Our results suggest that the PSF smoothing causes the measured bar ellipticity to be underestimated compared to its intrinsic value, and the underestimation becomes progressively more pronounced as the resolution decreases. The $\Delta{\epsilon}/\epsilon$ as a function of $n$ can be found in Fig.~\ref{fig:delvsN}.

Regarding the robustness in measuring the orientation of bars, we compare PA$_{\rm obs}^{n}$ with $\rm PA_{ bar,\,true}$ in Fig.~\ref{fig:pawithN}. Their mean difference ($\overline{{\Delta {\rm PA}}}$) and standard deviation ($\sigma_{\Delta {\rm PA}}$) are present in each panel. Since the absolute value of PA does not necessarily reflect physical significance, we have not normalized $\Delta{\rm PA}$ by ${\rm PA}_{\rm bar,\,true}$.  It's clear that, regardless of the resolution level, ${\rm PA}^n_{\rm obs}$ remains consistently close to its intrinsic value. The absolute value of $\overline{\Delta{\rm PA}}$ constantly stays less 1 degree. The $\Delta {\rm PA}$ as a function of $n$ can be found in Fig.~\ref{fig:delvsN}. Our results indicate that resolution has minimal influence on the measurement of the orientation of the bar.

Figure~\ref{fig:delvsN} plots the previously discussed change or fractional change ($\Delta{\epsilon}_{\rm obs}^{n}/\epsilon_{\rm bar,\,true}$, $\Delta{a}_{\rm obs}^{n}/a_{\rm bar,\,true}$, and $\Delta {\rm PA}^n_{\rm obs}$) between parameters measured from low-resolution images generated using the F200W filter and their intrinsic values as a function of resolution ($n=a_{\rm bar,\,true}/{\rm FWHM}$). These data, along with those obtained using F115W and F150W filters, are listed in Table~\ref{Correction table}. The dependence of $\Delta{\epsilon}^{n}_{\rm obs}/\epsilon_{\rm bar,\,true}$, $\Delta{a}^{n}_{\rm obs}/a_{\rm bar,\,true}$, and $\Delta {\rm PA}^n_{\rm obs}$ on $n$ for the F115W or F150W filter are quite similar to those for the F200W filter. However, as has been shown in Fig.~\ref{fig:BarfracN}, the $\eta_{\rm eff}$ for F115W filter is lower than those for F150W or F120W filter at a given $n>4$, since the F115W PSF is not Nyquist-sampled at a pixel scale of 0.03 arcsec/pixel. We note that there are fewer than 10 bars identified at $n=1.44$ and 1.2 for the F115W filter, at $n=1.2$ for the F150W filter, and at $n=1$ for the F200W filter, which might influence the statistical significance of the results at these $n$ values. 

\begin{table*}
\centering
\caption{
Biases and uncertainties in idenfifying and quantifying bars in the simulated low-resolution JWST images. Col. (1) resolution level $n = a_{\rm bar,\,true}/{\rm FWHM}$; (2) averaged value of the fractional difference ${{\Delta} a_{\rm obs}^{n} / a_{\rm bar,\,true}}$, where ${\Delta} a_{\rm obs}^{n} = (a_{\rm obs}^{n} - a_{\rm bar,\,true})$; (3) standard deviation of ${{\Delta} a_{\rm obs}^{n} / a_{\rm bar,\,true}}$; (4) averaged value of the fractional difference $\Delta{\epsilon_{\rm obs}^{n}}/\epsilon_{\rm bar,\,true}$, where ${\Delta} \epsilon_{\rm obs}^{n} = (\epsilon_{\rm obs}^{n} - \epsilon_{\rm bar,\,true})$; (5) standard deviation of ${{\Delta} \epsilon_{\rm obs}^{n} / \epsilon_{\rm bar,\,true}}$; (6) averaged value of the difference $\Delta{{\rm PA}_{\rm obs}^{n}} = {\rm PA}_{\rm obs}^{n}-{\rm PA}_{\rm bar,\,true}$; (7) standard deviation of ${\rm PA}_{\rm obs}^{n}$; (8) effectiveness of detecting bars at specific resolution level.} 

\label{Correction table}
\renewcommand\arraystretch{1.4}
\begin{tabular}{cccccccc}
\hline\hline 
\multicolumn{8}{c}{F115W}    
\\                                         
\hline      
$n $ & $\overline{\Delta{a_{\rm obs}^{n}}/a_{\rm bar,\,true}}$ 
& $\sigma_{\Delta{a_{\rm obs}^{n}}/a_{\rm bar,\,true}}$ & $\overline{\Delta{\epsilon_{\rm obs}^{n}}/\epsilon_{\rm bar,\,true}}$  & $\sigma_{\Delta{\epsilon_{\rm obs}^{n}}/\epsilon_{\rm bar,\,true}}$ & $\overline{\Delta{{\rm PA}_{\rm obs}^{n}}}$  & $\sigma_{\Delta{{\rm PA}_{\rm obs}^{n}}}$ & $\eta_{\rm eff}$ \\ 
{}&{}&{}&{}&{}&{($^{\circ}$)}&{($^{\circ}$)}&{} \\
(1)&(2)&(3)&(4)&(5)&(6)&(7)&(8)\\

\hline
10   & $-$0.03 & 0.08 & $-$0.07 & 0.06 & $-$0.12 & 2.58  & 1.00 \\
5.16 & $-$0.02 & 0.10 & $-$0.18 & 0.08 & $-$0.19 & 3.25  & 1.00 \\
4.3  & $-$0.02 & 0.11 & $-$0.22 & 0.08 & $-$0.16 & 4.02  & 0.99 \\
3.58 & $-$0.01 & 0.12 & $-$0.28 & 0.10 & $-$0.29 & 4.70  & 0.98 \\
2.99 & $-$0.02 & 0.14 & $-$0.34 & 0.11 & $-$0.48 & 5.78  & 0.89 \\
2.49 & $-$0.02 & 0.14 & $-$0.40 & 0.11 & $-$0.51 & 5.78  & 0.69 \\
2.07 & 0.03  & 0.14 & $-$0.42 & 0.12 & $-$0.53 & 6.46  & 0.35 \\
1.73 & 0.12  & 0.17 & $-$0.45 & 0.14 & 0.20 & 5.70  & 0.11 \\
1.44 & 0.11  & 0.00 & $-$0.57 & 0.02 & $-$8.33 & 8.92  & 0.01 \\
1.2  & $\cdots$    & $\cdots$  & $\cdots$  &$\cdots$  & $\cdots$ & $\cdots$ & 0.00 \\
1    & $\cdots$   & $\cdots$  & $\cdots$   & $\cdots$ & $\cdots$ & $\cdots$ & 0.00 \\
\hline\hline
\multicolumn{8}{c}{F150W}   
\\
\hline                                                                 
$n$  & $\overline{\Delta{a_{\rm obs}^{n}}/a_{\rm bar,\,true}}$ & $\sigma_{\Delta{a_{\rm obs}^{n}}/a_{\rm bar,\,true}}$ & $\overline{\Delta{\epsilon_{\rm obs}^{n}}/\epsilon_{\rm bar,\,true}}$ & $\sigma_{\Delta{\epsilon_{\rm obs}^{n}}/\epsilon_{\rm bar,\,true}}$ & $\overline{\Delta{{\rm PA}_{\rm obs}^{n}}}$ & $\sigma_{\Delta{{\rm PA}_{\rm obs}^{n}}}$ & $\eta_{\rm eff}$ \\ 
{}&{}&{}&{}&{}&{($^\circ$)}&{($^\circ$)}&{} \\
\hline
10   & $-$0.02 & 0.08 & $-$0.04 & 0.05 & $-$0.18 & 2.14  & 1.00 \\
5.16 & $-$0.03 & 0.10 & $-$0.14 & 0.08 & $-$0.17 & 2.46  & 1.00 \\
4.3  & $-$0.01 & 0.10 & $-$0.18 & 0.09 & $-$0.09 & 3.37  & 1.00 \\
3.58 & $-$0.02 & 0.11 & $-$0.23 & 0.10 & $-$0.13 & 4.06  & 0.99 \\
2.99 & $-$0.03 & 0.13 & $-$0.29 & 0.10 & $-$0.24 & 4.64  & 0.98 \\
2.49 & $-$0.04 & 0.12 & $-$0.34 & 0.10 & $-$0.33 & 4.98  & 0.94 \\
2.07 & $-$0.01 & 0.12 & $-$0.39 & 0.11 & $-$0.39 & 5.58  & 0.81 \\
1.73 & 0.03  & 0.13 & $-$0.44 & 0.12 & $-$0.69 & 6.27  & 0.53 \\
1.44 & 0.11  & 0.16 & $-$0.46 & 0.11 & 0.20 & 7.44  & 0.17 \\
1.2  & 0.15  & 0.14 & $-$0.51 & 0.12 & 2.73 & 5.53  & 0.02 \\
1    &$\cdots$    & $\cdots$   & $\cdots$ & $\cdots$  & $\cdots$ & $\cdots$ & 0.00 \\
\hline\hline
\multicolumn{8}{c}{F200W}
\\
\hline  
$n$  & $\overline{\Delta{a_{\rm obs}^{n}}/a_{\rm bar,\,true}}$ & $\sigma_{\Delta{a_{\rm obs}^{n}}/a_{\rm bar,\,true}}$ & $\overline{\Delta{\epsilon_{\rm obs}^{n}}/\epsilon_{\rm bar,\,true}}$ & $\sigma_{\Delta{\epsilon_{\rm obs}^{n}}/\epsilon_{\rm bar,\,true}}$ & $\overline{\Delta{{\rm PA}_{\rm obs}^{n}}}$ & $\sigma_{\Delta{{\rm PA}_{\rm obs}^{n}}}$ & $\eta_{\rm eff}$   \\ 
{}&{}&{}&{}&{}&{(${^\circ}$)}&{(${^\circ}$)}&{} \\
\hline
10   & $-$0.01 & 0.07 & $-$0.02 & 0.04 & 0.05 & 1.70 & 1.00 \\
5.16 & $-$0.01 & 0.09 & $-$0.10 & 0.08 & $-$0.20 & 2.32 & 1.00 \\
4.3  & $-$0.02 & 0.09 & $-$0.15 & 0.09 & $-$0.25 & 2.87 & 1.00 \\
3.58 & $-$0.02 & 0.12 & $-$0.20 & 0.10 & $-$0.27 & 3.79 & 1.00 \\
2.99 & $-$0.04 & 0.13 & $-$0.25 & 0.11 & $-$0.27 & 4.05 & 0.99 \\
2.49 & $-$0.06 & 0.11 & $-$0.30 & 0.10 & $-$0.34 & 4.23 & 0.95 \\
2.07 & $-$0.03 & 0.10 & $-$0.34 & 0.10 & $-$0.34 & 4.90 & 0.84 \\
1.73 & 0.02 & 0.13 & $-$0.40 & 0.11 & $-$0.43 & 6.37 & 0.59 \\
1.44 & 0.10 & 0.17 & $-$0.46 & 0.11 & 0.25 & 8.38 & 0.26 \\
1.2  & 0.17 & 0.18 & $-$0.49 & 0.10 & 0.98 & 7.38 & 0.06 \\
1    & 0.19 & 0.15 & $-$0.54 & 0.08 & 0.42 & 6.54 & 0.03 \\
\hline
\end{tabular}
\end{table*}


\begin{table*}
\centering
\caption{
Biases and uncertainties in identifying and quantifying bars in the simulated low-$S/N$ JWST images.   Col. (1) different $S/N$ level; (2) averaged value of the fractional difference ${{\Delta} a_{\rm obs}^{S/N} / a_{\rm obs}^{n=4.3}}$, where ${\Delta} a_{\rm obs}^{S/N} = (a_{\rm obs}^{S/N} - a_{\rm obs}^{n=4.3})$; (3) Standard deviation of ${{\Delta} a_{\rm obs}^{S/N} / a_{\rm obs}^{n=4.3}}$; (4) averaged value of the fractional difference $\Delta{\epsilon_{\rm obs}^{S/N}}/\epsilon_{\rm obs}^{n=4.3}$, where ${\Delta} \epsilon_{\rm obs}^{S/N} = (\epsilon_{\rm obs}^{S/N} - \epsilon_{\rm obs}^{n=4.3})$; (5) standard deviation of ${{\Delta} \epsilon_{\rm obs}^{S/N} / \epsilon_{\rm obs}^{n=4.3}}$; (6) averaged value of the difference $\Delta{{\rm PA}_{\rm obs}^{S/N}} = {\rm PA}_{\rm obs}^{S/N}-{\rm PA}_{\rm obs}^{n=4.3}$; (7) standard deviation of ${\rm PA}_{\rm obs}^{S/N}$; (8) effectiveness of detecting bars at specific $S/N$ level; (9) the effective surface brightness $\mu_e$. 
}
\label{SNR table2}
\renewcommand\arraystretch{1.4}
\begin{tabular}{ccccccccc}
\hline\hline
{$S/N$} & {$\overline{{\Delta} a_{\rm obs}^{S/N} / a_{\rm obs}^{n=4.3}}$} & {$\sigma_{{\Delta} a_{\rm obs}^{S/N} / a_{\rm obs}^{n=4.3}}$} & 

{$\overline{\Delta{\epsilon_{\rm obs}^{S/N}}/\epsilon_{\rm obs}^{n=4.3}}$} & {$\sigma_{\Delta{\epsilon_{\rm obs}^{S/N}}/\epsilon_{\rm obs}^{n=4.3}}$} & {$\overline{\Delta{{\rm PA}_{\rm obs}^{S/N}}}$} & {$\sigma_{\Delta{{\rm PA}_{\rm obs}^{S/N}}}$} & $\eta_{\rm eff}$ & $\overline{\mu_{e}}$ \\ 
{}&{}&{}&{}&{}&{($^{\circ}$)}&{($^{\circ}$)}&{} & {({mag arcsec$^{-2}$})}\\
(1)&(2)&(3)&(4)&(5)&(6)&(7)&(8) & (9)\\

\hline
62& 0.00& 0.02& 0.00& 0.05& $-0.02$& 0.02& 1.00& 18.96 \\
44.3 & 0.00 & 0.02 & 0.00 & 0.05 & 0.05 & 0.02 & 1.00 & 19.64  \\
31.6 & 0.00 & 0.02 & 0.00 & 0.05 & 0.03 & 0.02 & 1.00 & 20.30  \\
22.6 & 0.00 & 0.03 & 0.00 & 0.05 & $-0.08$ & 0.03 & 1.00 & 20.92  \\
16.1 & 0.00 & 0.02 & 0.01 & 0.05 & 0.04 & 0.02 & 1.00 & 21.51  \\
11.5 & 0.01 & 0.04 & 0.00 & 0.06 & $-0.01$ & 0.04 & 1.00 & 22.06  \\
8.2 & 0.01 & 0.03 & 0.01 & 0.06 & $-0.01$ & 0.03 & 1.00 & 22.57  \\
5.9 & 0.02 & 0.05 & 0.01 & 0.07 & $-0.15$ & 0.05 & 0.99 & 23.04  \\
4.2 & 0.03 & 0.05 & 0.01 & 0.08 & $-0.05$ & 0.05 & 0.99 & 23.50  \\
3 & 0.04 & 0.06 & 0.01 & 0.09 & $-0.46$ & 0.06 & 0.96 & 23.93 \\
\hline
\end{tabular}
\end{table*}

\subsection{The effect of noise}\label{results of SNR}
We investigate the influence of noise on the bar analysis by comparing the parameters ($a_{\rm obs}^{S/N}$, $\epsilon_{\rm obs}^{S/N}$, and ${\rm PA}_{\rm obs}^{S/N}$), measured in the simulated low-$S/N$ images, and their noiseless values ($a_{\rm obs}^{n=4.3}$, $\epsilon_{\rm obs}^{n=4.3}$, and ${\rm PA}_{\rm obs}^{n=4.3}$), measured the noiseless low-resolution simulated images of $n=4.3$. The $n$ is fixed to 4.3 for isolating the effects from noise. As These measurements are relatively robust against noise, we refrain from showing the plots, but list the results in Table~\ref{SNR table2}. This table provides $\eta_{\rm eff}$ (Eq.~[\ref{eff}]), and the mean value ($\overline{\Delta{a_{\rm obs}^{S/N}}/a_{\rm obs}^{n=4.3}}$, $\overline{\Delta{\epsilon_{\rm obs}^{S/N}}/\epsilon_{\rm obs}^{n=4.3}}$, and $\overline{\Delta{{\rm PA}_{\rm obs}^{S/N}}}$) and standard deviation ($\sigma_{\Delta{a_{\rm obs}^{S/N}}/a_{\rm obs}^{n=4.3}}$, $\sigma_{\Delta{\epsilon_{\rm obs}^{S/N}}/\epsilon_{\rm obs}^{n=4.3}}$, and $\sigma_{\Delta{\rm PA}_ {\rm obs}^{S/N}}$) of the change or fractional change of the measured parameters, defined in the same vein as previously did. It can be seen from Table~\ref{SNR table2} that, when $S/N \geq 11.5$, the $\eta_{\rm eff}$ remains 100\%, suggesting that within this $S/N$ range, all the barred galaxies are able to be detected. As the $S/N$ falls below 11.5, there are a tiny fraction of barred galaxies can be missed. Nevertheless, even though when $S/N$ reaches as low as 3, almost all the bars can be identified. Irrespective of the $S/N$ level, the deviation of the measured $a_{\rm obs}^{S/N}$, $\epsilon_{\rm obs}^{S/N}$ and PA$_{\rm obs}^{S/N}$ of bars from their noiseless values are quite small, most of which are $\leq 5\%$ for $a$, $\leq 1\%$ for $\epsilon$, and $\leq 0.5$ degree for PA, though the scatter slightly increase with lower $S/N$.  The above analysis has also been performed for the noiseless simulated images of other values of $a_{\rm bar,\,true}/{\rm FWHM}$, and the results are almost identical to Table~\ref{SNR table2}. Our results suggest that, under the typical $S/N$ range in CEERS for galaxies with redshifts of $z\leq 3$ and stellar mass of $M_{\star} \geq 10^{9.75} M_{\odot}$, the detection and quantification of bars are not significantly adversely affected by the noise.

In practical applications, it is more straightforward to calculate the effective surface brightness ($\mu_e$), which has similar ability to characterize the clarity of an extended structure similarly to the $S/N$ for a given background noise. To make our results more easily utilized, we compute the mean value of $\mu_e$ ($\overline{\mu_e}$) for our simulated images at each $S/N$ level and list them in Table~\ref{SNR table2}.  As our simulated images are generated to match the typical noise conditions in the CEERS field, the values of $\overline{\mu_e}$ only work robustly in the CEERS field. These $\overline{\mu_e}$, corresponding to a given $S/N$, should be fainter in a survey deeper than CEERS but brighter in a shallower survey.

\begin{figure}
\begin{center}
\centering
\includegraphics[width=0.45\textwidth]{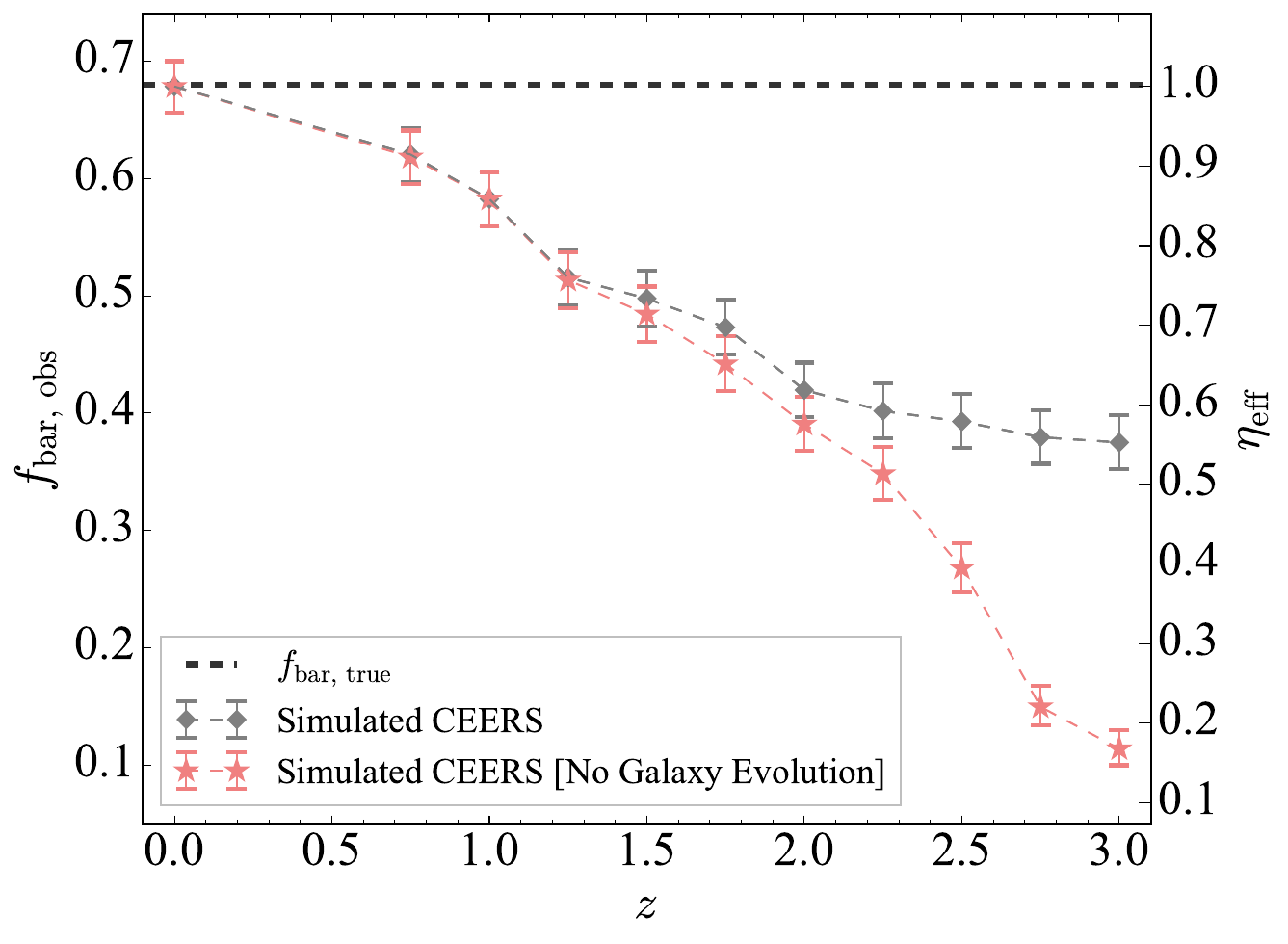}
\caption{Dependence of fraction of bars ($f_{\rm bar,\,obs}$) and effectiveness of detecting bars ($\eta_{\rm eff}$), measured from the simulated CEERS images, as a function of redshift ($z$). The grey diamonds represent the results obtained from the simulated data with the galaxy evolution model, while the pink stars represent those from the data without. Error bars denote the uncertainty of $f_{\rm bar,\,obs}$. For the simulated CEERS images, the F115W, F150W, and F200W filters are used for redshift range $z=0.75$\textendash1.0, 1.25\textendash1.75, and 2.0\textendash3.0, respectively. The horizontal dashed line represents the $f_{\rm bar,\,true}$ of 68\% of our sample.
}
\label{fig:Bar_frac_zhi}
\end{center}
\end{figure}

\subsection{Bar identification and measurement in simulated CEERS images}\label{results of CEERS}

In this section, we explore how the identification and quantification of bars at high redshifts are influenced by the combination of observational effects and evolution effects using the simulated CEERS images with galaxy evolution models included. The dependence of method effectiveness of detecting bars ($\eta_{\rm eff}$; on the left y-axis) and observed bar fraction ($f_{\rm bar,\,obs}$; on the right y-axis) on redshift $z$ are shown by the grey diamonds  in Fig.~\ref{fig:Bar_frac_zhi}. The error bars associated with the data points are the uncertainty of $f_{\rm bar,\,obs}$. The $f_{\rm bar,\,obs}$ is defined as the ratio of number of bars identified at each redshift to the total number of disk galaxies:
\begin{equation}\label{fbarobs}
  f_{\rm bar,\,obs}=\frac{\rm Number~of~detected~bars}{\rm Number~of~disks}.
\end{equation}
\noindent
As redshift increases from 0 to 3, the $\eta_{\rm eff}$ exhibits a gradual decrease, declining from 100\% to approximately 55\%, suggesting more and more bars being missing at higher redshifts due to the impact of observational effects. Approximately half of the bars can go undetected at $z=3$. Meanwhile, the $f_{\rm bar,\,obs}$ decreases from 68\% to 37.5\%, suggesting that the bar fraction observed at JWST F200W band can be underestimated by 30\% at $z=3$ compared to local universe. The $\eta_{\rm eff}$-$z$ and $f_{\rm bar,\,obs}$-$z$ relations are not smooth but present two sudden declines at $z=1.0$ and $z=1.75$, which stems from the changes of the filters. From $z=1.0$ to $z=1.25$, for better tracing the rest-frame optical light, the F115W is changed to F150W, the image PSFs broadens, causing the images to become blurrier and thus reducing the bar detection capability. When the F150W filter is changed to F200W at $z=2.0$, the same principle applies. The trend of decreasing of apparent $f_{\rm bar,\,obs}$ with increasing $z$ is consistent with \cite{Erwin2018}, who used S$^4$G images to simulate the resolution of high-redshift HST images and generated a decreasing trend of bar fraction toward higher redshifts by applying a cut in bar size to select barred galaxies.

For comparison, we shown the results for the simulated images without involving galaxies evolutions as pink stars in Fig.~\ref{fig:Bar_frac_zhi}. The measured $f_{\rm bar,\,obs}$ at $z\leq1.25$ is almost identical to the results from the simulated CEERS image with involving galaxies evolutions. However, as redshift increases, the $f_{\rm bar,\,obs}$ decreases significantly, reaching 11\% at $z=3.0$. This decrease is caused by the $S/N$ being too low due to the cosmological dimming effect, as discussed in Sect.~\ref{sect:sim}.

\begin{figure}
\begin{center}
\centering
\includegraphics[width=0.45\textwidth]{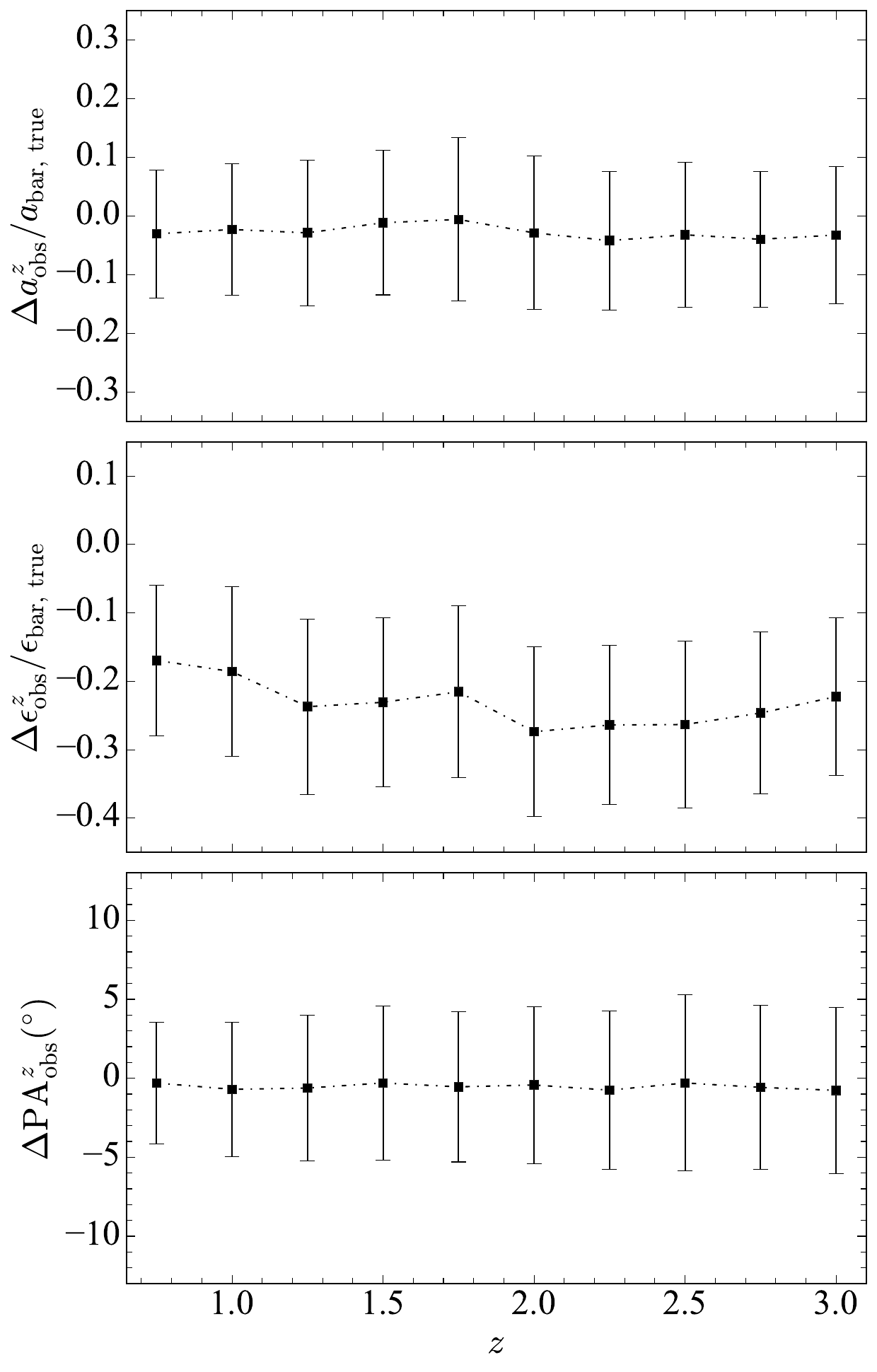}
\caption{Fractional difference ($\Delta a^z_{\rm obs}/a_{\rm bar,\,true}$ and $\Delta\epsilon^z_{\rm obs}/\epsilon_{\rm bar,\,true}$) or absolute difference ($\Delta$PA$^z_{\rm obs}$) between parameters measured from the simulated CEERS images and their intrinsic values plotted against redshift ($z$).
The error bar represents the standard deviation of the results at each redshift. 
} 
\label{fig:delvszhi}
\end{center}
\end{figure}

In Fig.~\ref{fig:delvszhi}, we plot the fractional difference ($\Delta a_{\rm obs}^z/a_{\rm bar,\,true}$, $\Delta\epsilon_{\rm obs}^z/\epsilon_{\rm bar,\,true}$) or absolute difference ($\Delta{\rm PA}^z_{\rm obs}$) between the three parameters measured from the simulated CEERS images at each redshift and their intrinsic values. As shown in the top row, the ${\Delta} a^{z}_{\rm obs}/a_{\rm bar,\,true}$ at each redshift is slightly less than zero on average, suggesting that the bar size measurements at higher redshifts are relatively robust, with only a few percent underestimation ($\leq 4\%$) to be noted. The relation between ${\Delta} a^{z}_{\rm obs}/a_{\rm bar,\,true}$ and $z$ does not reveal the systematic overestimation in measured bar size when resolution is very low as might be expected from Fig.~\ref{fig:delvsN}. This is because most short bars with size $a_{\rm bar,\,true} \lesssim 2\times {\rm FWHM}$ have been missed (Fig.~\ref{fig:BarfracN}) in the process of bar identification for the simulated CEERS images. Specifically, at $z=3$, there are 90\% of identified bars with size $a_{\rm bar,\,true} > 2\times {\rm FWHM}$, therefore leading to a minor underestimation of measured bar size according to Fig.~\ref{fig:delvsN}. The relation between $\Delta\epsilon^{z}_{\rm obs}/\epsilon_{\rm bar,\,true}$ and $z$ shows that the measured bar ellipticities are on average more and more underestimated at redshift increasing from $z=0.75$ to $z=2.0$, but the underestimation become approximately constant beyond $z=2.0$. One might think that, as galaxies become angularly smaller at high redshifts, the $\Delta\epsilon^{z}_{\rm obs}/\epsilon_{\rm bar,\,true}$ should decrease strictly monotonically with higher $z$ according to Fig.~\ref{fig:delvsN}, but this conjecture is not true. The decline in $\Delta\epsilon^{z}_{\rm obs}/\epsilon_{\rm bar,\,true}$ from $z=0.75$ to $z=2$ is indeed due to the significant reduction in the angular size of galaxies, in which effects from longer distance and size evolution of the disks are considered. However, at higher redshifts, more and more bars become angularly too small to be detected, so that the contribution of severely underestimation in $\epsilon$ caused by short bars to the calculation of mean value is significantly reduced, leading to a approximate constant average $\Delta\epsilon^{z}_{\rm obs}/\epsilon_{\rm bar,\,true}$ between $z=2$ and $z=3$. As expected, the measurement of PA is quite robust without any obvious systematic biases.

\subsection{Correction of the measurement biases}\label{Correction of resoluton}

\begin{figure}
\begin{center}
\centering
\includegraphics[width=0.45\textwidth]{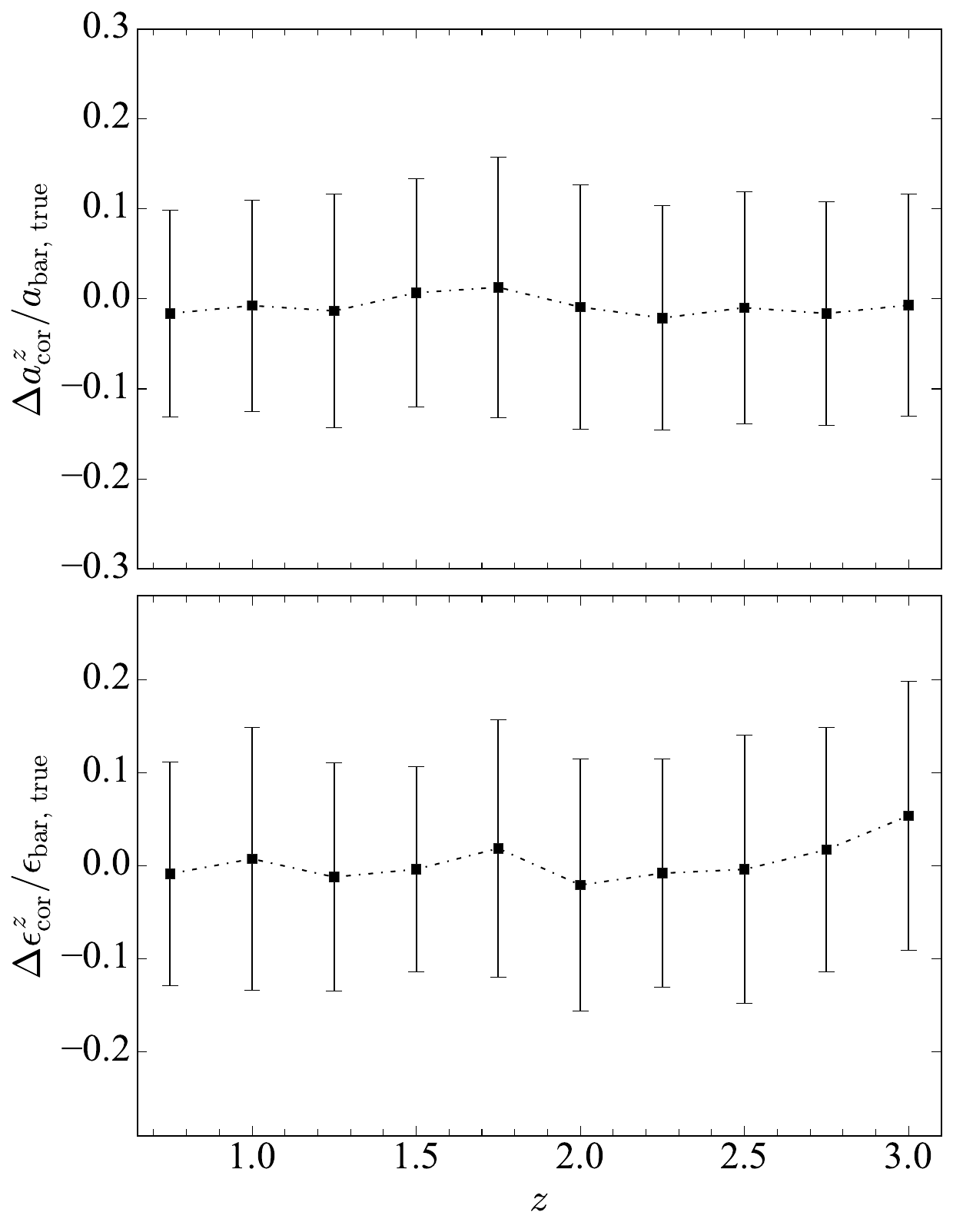}
\caption{
Fractional difference ($\Delta a_{\rm cor}^z/a_{\rm bar,\,true}$ and $\Delta\epsilon_{\rm cor}^z/\epsilon_{\rm bar,\,true}$) between bias-corrected parameters measured from the simulated CEERS images and their intrinsic values plotted against the redshift ($z$). The error bar represents the standard deviation of the results at a given redshift. 
} 
\label{fig:Cor_delvszhi}
\end{center}
\end{figure}

While the PA measurement remains unbiased, the measurements of $a$ and $\epsilon$ are biased towards lower values at high redshifts, with resolution effects being the primary cause. With the fractional differences between the measurements from the low-resolution images and their intrinsic values listed in Table~\ref{Correction table} , we can effectively remove the biases introduced by resolution. The biased-corrected bar size, denoted by $a_{\rm cor}^z$, can be obtained through
\begin{equation} \label{corr equation r}
a_{\rm cor}^z = \frac{a^{z}_{\rm obs}}{(\overline{\Delta a_{\rm obs}^{n}/a_{\rm bar,\,true}}+1)},
\end{equation}
where $n$ is approximated as $a^{z}_{\rm obs}/\rm FWHM$ and the value of $\overline{\Delta a_{\rm obs}^{n}/a_{\rm bar,\,true}}$ is obtained via interpolation based on the data in Table~\ref{Correction table}. The same methodology can be applied to obtain the biased-corrected bar ellipticity, denoted as $\epsilon_{\rm cor}^z$:
\begin{equation} \label{corr equation e}
\epsilon_{\rm cor}^z = \frac{\epsilon^{z}_{\rm obs}}{(\overline{\Delta\epsilon_{\rm obs}^{n}/\epsilon_{\rm bar,\,true}}+1)}, 
\end{equation}
where the value of $\overline{\Delta\epsilon_{\rm obs}^{n}/\epsilon_{\rm bar,\,true}}$ is obtained through interpolation. Figure~\ref{fig:Cor_delvszhi} plots the fractional difference ($\Delta a_{\rm cor}^z/a_{\rm bar,\,true}$) between $a_{\rm cor}^z$ and its intrinsic value $a_{\rm bar,\,true}$ in the top panel and the fractional difference ($\Delta\epsilon_{\rm cor}^z/\epsilon_{\rm bar,\,true}$) between $\epsilon_{\rm cor}^z$ and its intrinsic value $\epsilon_{\rm bar,\,true}$ in the bottom panel. After making corrections, the values of $a_{\rm cor}^z$ and $\epsilon_{\rm cor}^z$ underestimate $a_{\rm bar,\,true}$ and  $\epsilon_{\rm bar,\,true}$ by less than 1\% on average. The residuals are negligibly small, indicating that our correction functions are effective.

\begin{table*}[]
\centering
\caption{Summary of bar fractions measured from simulated CEERS images at different redshifts using different methods. Col. (1) redshift; (2) the bar fraction ($f_{\rm bar,\,obs}$) measured from simulated CEERS images using the ellipse fitting method as described in Sect~\ref{sect:obs}; (3)--(6) the $f_{\rm bar,\,obs}$ estimated from simulated CEERS images by adopting the cut of $a_{\rm bar,\,true} > 2\times {\rm FWHM}$ for bar detection for four different NIRcam filters.}
\label{Frac table}
\renewcommand\arraystretch{1.4}
\begin{tabular}{cccccc}

\hline\hline
$z$  & $f_{\rm bar,\,obs}$(ellipse)  &  $f_{\rm bar,\,obs}^{\rm F200W}$(cut)  & $f_{\rm bar,\,obs}^{\rm F277W}$(cut) & $f_{\rm bar,\,obs}^{\rm F356W}$(cut)  & $f_{\rm bar,\,obs}^{\rm F444W}$(cut)  \\
(1)&(2)&(3)&(4)&(5)&(6)\\
\hline
0.0  & 0.68 & 0.68  & 0.68  & 0.68  & 0.68  \\
0.75 & 0.62 & 0.55  & 0.46  & 0.36  & 0.27  \\
1.0  & 0.58 & 0.51  & 0.40  & 0.29  & 0.19  \\
1.25 & 0.52 & 0.47  & 0.35  & 0.23  & 0.15  \\
1.5  & 0.50 & 0.45  & 0.32  & 0.19  & 0.13  \\
1.75 & 0.47 & 0.42  & 0.29  & 0.17  & 0.10  \\
2.0  & 0.42 & 0.40  & 0.27  & 0.15  & 0.08  \\
2.25 & 0.40 & 0.39  & 0.25  & 0.15  & 0.08  \\
2.5  & 0.39 & 0.38  & 0.23  & 0.14  & 0.06  \\
2.75 & 0.38 & 0.37  & 0.22  & 0.13  & 0.06  \\
3.0  & 0.38 & 0.37  & 0.21  & 0.12  & 0.06 \\
\hline
\end{tabular}
\end{table*}

\section{Implication}\label{sect:implication}
Determining the bar fraction at high redshifts is crucial to understand the nature of bars, but it is challenging due to band shifting and/or image degradation \citep{Sheth2008}. Early HST-based study of the bar fraction evolution reported a significant decrease beyond $z\sim 0.5$ \citep{Abraham1999}. Then \cite{Sheth2003} found that the fraction of strong bars at $z>0.7$ (4/95) is higher than the fraction observed at $z<0.7$ (1/44). These fractions are likely lower limits, primarily due to resolution effects and the small-number statistic. With the images of improved resolution obtained from the HST Advanced Camera for Surveys (ACS), two subsequent studies of \cite{Elmegreen2004} and \cite{Jogee2004} reported a relatively consistent fraction up to redshifts $z\sim1$. Nevertheless, their samples are still of a modest size. With a statistical large sample of more than 2000 galaxies defined from the Cosmic Evolution Survey \citep[COSMOS;][]{Scoville2007}, \cite{Sheth2008} showed that the bar fraction declines from $\sim$\,65\% in the local universe to $\sim$\,20\% at $z\approx 0.84$. Although the overall bar fractions calculated based on Galaxy Zoo  have been underestimated \citep{Erwin2018}, studies based on it have consistently identified a trend declining from $\sim$ 22\% at $z=0.4$ to $\sim$ 11\% at $z=1.0$ \citep{Melvin2014}.

\begin{figure}
\begin{center}
\centering
\includegraphics[width=0.48\textwidth]{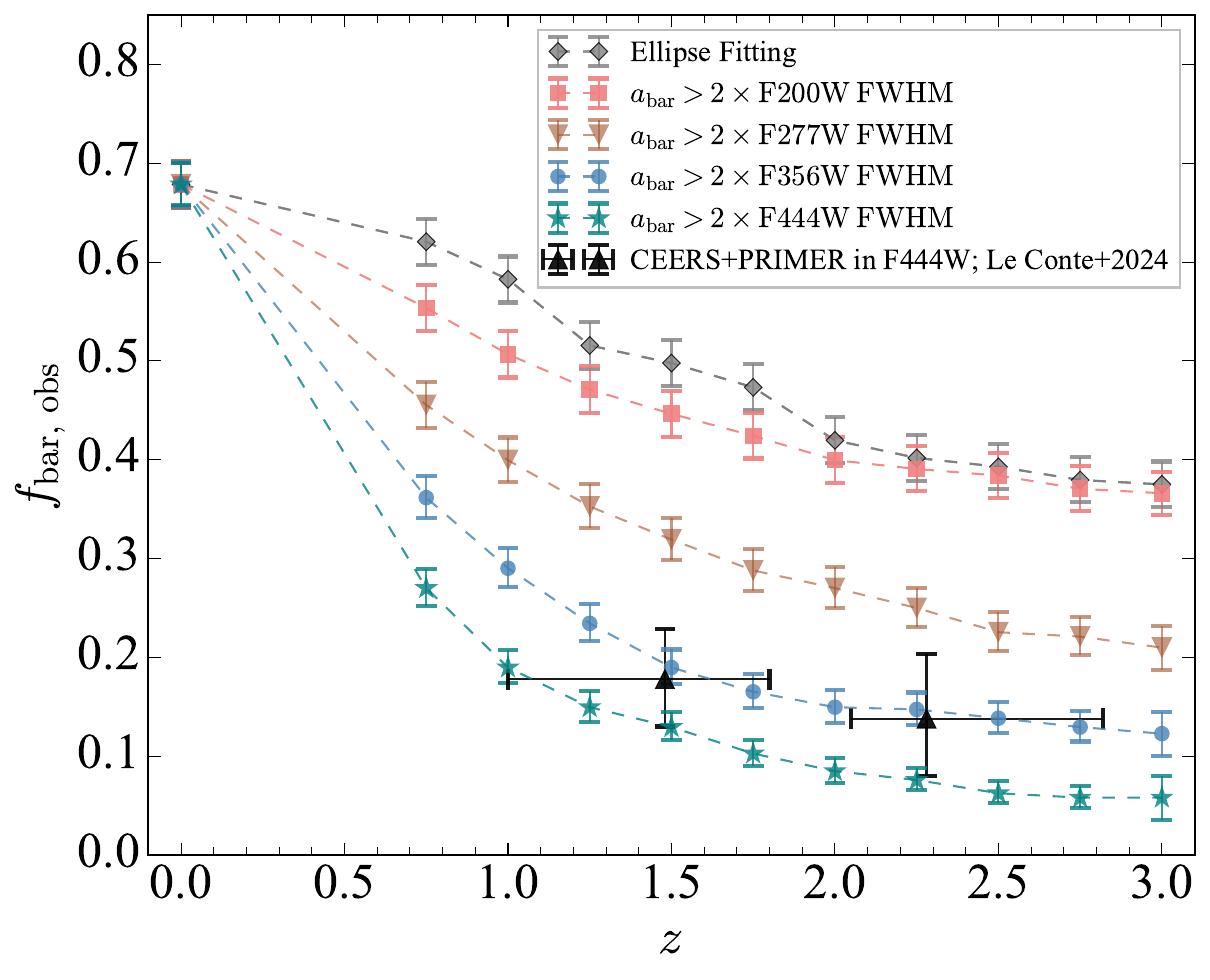}
\caption{
The measured bar fraction ($f_{\rm bar, obs}$) obtained from simulated CEERS images as a function of redshifts ($z$).  The grey diamonds represent the $f_{\rm bar, obs}$ obtained using the ellipse fitting method for simulated CEERS images. The other symbols mark the results obtained by adopting the criterion of $a_\text{ bar, true} > 2\times {\rm FWHM}$ for bar detection for specific JWST NIRcam filters.  The pink rectangles, brown inverted triangles, blue dots, and green stars represent the $f_{\rm bar, obs}$ for F200W, F277W, F356W, and F444W filter, respectively.
Black triangles correspond to the observed F444W-band bar fraction measured by \cite{LeConte2024}.
} 
\label{fig:cut_2FWHM}
\end{center}
\end{figure}

Previous studies have consistently raised concerns about the possibility of missing short bars at higher redshifts due to resolution limitations. Lacking rigorous quantitative justification, \cite{Sheth2003} proposed a bar size threshold of 2.5 times the PSF FWHM for bar detection. Concluding from experiments on artificial galaxies, \cite{Aguerri2009} proposed the same the threshold. \cite{Erwin2018} suggested a smaller threshold of 2 times the PSF FWHM. In Fig.~\ref{fig:BarfracN}, we demonstrate that, for images with a Nyquist-sampling PSF, the effectiveness of detecting bars $\eta_{\rm eff}$ remains at $\sim$\,100\% when $a_{\rm bar,\,true}/{\rm FWHM}$ is above 2. When $a_{\rm bar,\,true}/{\rm FWHM}$ is below 2, the $\eta_{\rm eff}$ declines sharply.  We select the bars in the simulated CEERS images using criterion $a_{\rm bar,\,true}>{\rm F200W~FWHM}$, calculate the $f_{\rm bar,\,obs}$, and plot them as pink square in Fig.~\ref{fig:cut_2FWHM}. These $f_{\rm bar,\,obs}$ are fully consistent with the F200W-band $f_{\rm bar,\,obs}$ derived from the ellipse fitting method, which are represented as grey diamonds in the plot. This suggests that using $a_{\rm bar,\,true}=2\times {\rm FWHM}$ as the bar detection threshold provides a better fit to the results obtained through the ellipse fitting method. Using the factor of 2.5 would underestimate the bar-detection efﬁciency of the ellipse fitting method. Nevertheless, if the PSFs are not Nyquist-sampling, the effectiveness of detecting bars starts to decline at higher $a_{\rm bar,\,true}/{\rm FWHM}$, suggesting a higher bar size threshold. The concern of missing bars due to resolution effects is further amplified by the growth of bar size over cosmic time, as indicated by simulation studies \citep[e.g.,][]{Debattista2000,Martinez2006,Algorry2017,Rosas-Guevara2022}.

Meanwhile, studies on bars based on cosmological simulations also highlight the potential impacts rising from resolution effects. By studying TNG100 galaxies, \cite{Zhao2020} revealed a roughly constant bar fraction of $\sim$ 60\% at $0<z<1$ when selecting galaxies with a mass cut of $M_{\star}\geq 10^{10.6} M_{\odot}$. However, considering the resolution limitations observed in HST images, where bars shorter than 2 kpc can be missed at z $\sim 1$, they focus on bars longer than 2.2 kpc and consequently detect a decreasing trend of bar fraction at higher redshifts. Moreover, \cite{Rosas-Guevara2022} used the data from TNG50 to study spiral galaxies with $M_{\star}\geq10^{10}M_{\odot}$, showing that the derived bar fraction increases from 28\% at $z=4$, and reaches a peak of 48\% at $z=1$ and then drops to 30\% at $z=0$. Considering the $2\times\rm FWHM$ detection limit, they implemented an angular resolution limit equivalent to twice the HST F814W PSF FWHM. As a result, the bar fraction exhibited a decrease with increasing redshift at $z>0.5$, which relatively reconciles the differences between their results and observations.

With the advent of JWST, obtaining deep high-resolution NIR images has become accessible, enabling us to explore the structures in high-redshift galaxies in detail. Recently, \cite{Guo2023} analyse rest-frame NIR galaxy structures using F444W images from JWST CEERS and report the detection of six strongly barred galaxies at $1<z<3$.

\cite{Costantin2023} remarkably reported a barred galaxy at $z\simeq 3$ from CEERS, making it the most distant barred galaxy ever detected. This discovery suggests that dynamically cold disk galaxies could have already been in place at $z=4\text{\textendash}5$. Our findings, derived from simulated CEERS data, can serve as a comparison sample for studies of high-redshift bar fraction based on JWST. Grey diamonds in Figure~\ref{fig:cut_2FWHM} plots the observed $f_{\rm bar,\,obs}$ from the simulated CEERS images in the F115W band for $z=0.75$\textendash1.0, F150W band for $z=1.25$\textendash1.75, and F200W band for $z=2.0$\textendash3.0. We restate that the CEERS image simulation procedure incorporates the changes in angular size due to both distance and the intrinsic evolution of disk size. As a result, bars in the simulated CEERS images appear shorter in physical size at higher redshift, directly following the disk size evolution described by \cite{Vanderwel2014}. Specifically, at redshifts $z=1$, $z=2$, and $z=3$, bars are approximately 63\%, 48\%, and 40\% the size of their local counterparts. In contrast, the ratio of the bar size to disk size remains unchanged, and the true bar fraction is fixed to 68\%, the value measured in DESI images (Sect.~\ref{sect:obs}). To present the results based on other long-wavelength NIRcam filters (F277W, F356W, and F444W), we estimate the measured $f_{\rm bar,\,obs}$ using the current simulated dataset by adopting the criterion of $a_{\rm bar,\,true} > 2\times {\rm FWHM}$ for bar detection, as proposed by \cite{Erwin2018}.  First, we apply $a_{\rm bar,\,true}>2\times {\rm F200W~FWHM}$ to detect bars, calculate $f_{\rm bar,\,obs}$ observed in the F200W band, and plot them as pink squares in Fig.~\ref{fig:cut_2FWHM}. They are in good agreement with the results at redshift $2\leq z\leq 3$ obtained from the ellipse fitting method, demonstrating the effectiveness of the $a_{\rm bar,\,true}>2\times {\rm F200W~FWHM}$ criterion. Subsequently, using $a_{\rm bar,\,true}>2\times {\rm F277W~FWHM}$, we calculate the F277W-band $f_{\rm bar,\,obs}$ observed in the simulated CEERS images, and plot them as green stars. The observed $f_{\rm bar,\,obs}$ in the F277W band is lower than its F200W-band counterpart because the F277W PSF FWHM is larger than the F200W PSF FWHM. The same calculations are applied to the F356W and F444W bands. We summarize the measured values of $f_{\rm bar,\,obs}$ as a function of redshifts for the four NIRcam bands in Table~\ref{Frac table}.

We note that the F444W-band $f_{\rm bar,\,obs}$ decreases significantly from 68\% at $z\approx0$ to 13\% at $z=1.5$, and further to 6\% at $z=3.0$. We plot the F444W-band bar fraction observed in JWST CEERS and the Public Release Imaging for Extragalactic Research \citep[PRIMER;][]{Dunlop2021} measured by \cite{LeConte2024} as black triangles in Fig.~\ref{fig:cut_2FWHM}. 
The F444W-band $f_{\rm bar,\,obs}$ from the simulated CEERS images is consistent within 1\,$\sigma$ uncertainty with those obtained from  JWST F444W-band observations reported in \cite{LeConte2024}, despite potential difference in the sample properties. Therefore, by accounting for resolution effects and bar size evolution, we have successfully largely reproduced the bar fraction observed by JWST in the F444W band without including evolution of intrinsic bar fraction. Our findings are consistent with \cite{Erwin2018}, who similarly factored in resolution effects and assumed bar sizes to be half their actual values at high redshifts. Using S$^4$G images for simulated observations, they reproduced the relation between observed bar fraction and stellar mass at redshifts up to 0.84, as reported by \cite{Sheth2008}. Our results imply that by positing the presence of all local bars as early as $z\sim3$, the combination of resolution effects and bar size growth can largely account for the apparent redshift evolution in observed bar fraction found by \cite{LeConte2024}.  Any intrinsic evolution in bar fraction, if it exists, might be artificially exaggerated by these factors. To truly grasp the evolution of intrinsic bar fraction, it's imperative to disentangle it from resolution effects and bar size evolution.

\section{Conclusions}\label{sect:conclusions}
Quantifying the evolution of bar fraction and bar properties is essential for understanding the evolutionary history of disk galaxies. During the HST era, a series of studies have extensively explored this subject \citep[e.g.,][]{Elmegreen2004,Jogee2004,Sheth2008,Perez2012,Melvin2014,Kim2021}. However, these results are potentially affected by limitations of image quality. Nowadays, with the superior high-resolution and deep NIR imaging available from JWST, bars in galaxies at high redshift can be studied in far more detail and some bars at $z>2$ have been successfully detected \citep{Guo2023, LeConte2024}. But still, the difficulties caused by limited observations are unavoidable, making it challenging to fully embrace the intrinsic results.   To assess our ability to analyze bars in high-redshift galaxies observed by JWST, we use a sample of 448 nearby face-on spiral galaxies, a subset of the sample conducted by \cite{Yu2023}, to simulate images under various observational conditions consistent with the JWST CEERS field, identify bars and quantify bar properties, and then compare the results before and after simulation to determine the systematic biases arising from resolution, noise, or the combination of both. The intrinsic bar size is denoted as $a_{\rm bar,\,true}$. The ratio $a_{\rm bar,\,true}/{\rm FWHM}$ is used to gauge the detectability of bars. Our main findings can be summarized as follows:

\begin{enumerate}
\item Both the identification and quantification of bars are hardly affected by noise when the $S/N$ is greater than 3, an observational condition met by CEERS galaxies with $M_{\star} \geq 10^{9.75} M_{\odot}$ at $z<3$. \\ 

\item For the F200W PSF, which is Nyquist-sampled, the effectiveness of detecting bars remains at $\sim$\,1 when $a_{\rm bar,\,true}/{\rm FWHM}$ is above a critical value of 2; when $a_{\rm bar,\,true}/{\rm FWHM}$ is below 2, the effectiveness drops sharply. The fractions of bars determined through ellipse fitting method is in good agreement with that derived using the criterion $a_{\rm bar,\,true} >2\times{\rm FWHM}$, a bar size threshold suggested by \cite{Erwin2018} for bar detection. Nevertheless, when the PSF is sub-Nyquist-sampled, the critical $a_{\rm bar,\,true}/{\rm FWHM}$ increases. For instance, For the F115W PSF at a pixel scale of 0.03 arcsec/pixel, this critical value is $\sim$\,3. \\

\item By assuming all local bars were already in place at high redshifts, we show that a combination of resolution effects and bar size growth can explain the apparent evolution of bar fraction obtained from JWST observations reported by \cite{LeConte2024}. This implies that the reported bar fraction has been significantly underestimated. The true bar fraction evolution, if it exists, could be shallower than detected. Our results underscores the importance of disentangling the true bar fraction evolution from resolution effects and bar size growth. \\

\item The measured bar size and bar ellipticity are typically underestimated, with the extent depending on $a_{\rm bar,\,true}/{\rm FWHM}$. In contrast, the measurement of the bar position angle remains unaffected by resolution. To remove these resolution effects, we have developed correction functions. When applied to the bar properties measured from the simulated CEERS images of high-redshift galaxies, these corrections yield bias-corrected values closely matching their intrinsic values.

\end{enumerate}

\begin{acknowledgements}
This work was supported by the National Science Foundation of China (11721303, 11890692, 11991052, 12011540375, 12133008, 12221003, 12233001), the National Key R\&D Program of China (2017YFA0402600, 2022YFF0503401), and the China Manned Space Project (CMS-CSST-2021-A04, CMS-CSST-2021-A06). We thank the referee for his insightful and constructive feedback, which significantly enhanced the quality and clarity of this letter. SYY acknowledges the support by the Alexander von Humboldt Foundation.  Kavli IPMU is supported by World Premier International Research Center Initiative (WPI), MEXT, Japan

\end{acknowledgements}

\bibliographystyle{aa}


\end{document}